\documentclass[iop, twocolumn]{aastex6}
\hypersetup{citecolor=blue,urlcolor=red}
\usepackage{amsmath}
\usepackage{url}
\usepackage{enumerate}
\usepackage{multirow}
\usepackage{caption}
\usepackage{threeparttable}

\newcommand{\h}{^{\rm h}}
\newcommand{\m}{^{\rm m}}
\newcommand{\s}{^{\rm s}}

\newcommand{\Trh}{t_{\rm rh}}
\newcommand{\Tdis}{t_{\rm d}}
\newcommand{\Tcr}{t_{\rm cr}}
\newcommand{\rh}{r_{\rm h}}
\newcommand{\rt}{r_{\rm t}}

\usepackage{lineno}

\begin{document}
\title{Discovery and Timing of Millisecond Pulsars in the Globular Cluster M5 with FAST and Arecibo}

\author{
Lei Zhang$^{1*,2}$\href{https://orcid.org/0000-0001-8539-4237}{\includegraphics[scale=0.08]{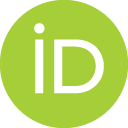}},
Paulo C. C. Freire$^{3}$\href{https://orcid.org/0000-0003-1307-9435}{\includegraphics[scale=0.08]{ORCIDiD.png}},
Alessandro Ridolfi$^{4,3}$\href{https://orcid.org/0000-0001-6762-2638}{\includegraphics[scale=0.08]{ORCIDiD.png}}, 
Zhichen Pan$^{1,5,6,7}$\href{https://orcid.org/0000-0001-7771-2684}{\includegraphics[scale=0.08]{ORCIDiD.png}},
Jiaqi Zhao$^{8}$\href{https://orcid.org/0000-0002-7716-1166}{\includegraphics[scale=0.08]{ORCIDiD.png}},
Craig O. Heinke$^{8}$\href{https://orcid.org/0000-0003-3944-6109}{\includegraphics[scale=0.08]{ORCIDiD.png}},
Jianxing Chen$^{9,10}$\href{https://orcid.org/0000-0002-8004-549X}{\includegraphics[scale=0.08]{ORCIDiD.png}},
Mario Cadelano$^{9,10}$\href{https://orcid.org/0000-0002-5038-3914}{\includegraphics[scale=0.08]{ORCIDiD.png}},
Cristina Pallanca$^{9,10}$\href{https://orcid.org/0000-0002-7104-2107}{\includegraphics[scale=0.08]{ORCIDiD.png}},
Xian Hou$^{11,12,13}$\href{https://orcid.org/0000-0003-0933-6101}{\includegraphics[scale=0.08]{ORCIDiD.png}},
Xiaoting Fu$^{14,10}$\href{https://orcid.org/0000-0002-6506-1985}{\includegraphics[scale=0.08]{ORCIDiD.png}}, 
Shi Dai$^{15,1}$\href{https://orcid.org/0000-0002-9618-2499}{\includegraphics[scale=0.08]{ORCIDiD.png}}, 
Erbil G\"{u}gercino\u{g}lu$^{1}$,
Meng Guo$^{16,17}$,
Jason Hessels$^{18,19}$\href{https://orcid.org/0000-0003-2317-1446}{\includegraphics[scale=0.08]{ORCIDiD.png}},
Jiale Hu$^{17}$,
Guodong Li$^{1}$\href{https://orcid.org/0000-0003-4007-5771}{\includegraphics[scale=0.08]{ORCIDiD.png}},
Mengmeng Ni$^{17}$,
Jingshan Pan$^{16,17}$\href{https://orcid.org/0009-0002-0968-0658}{\includegraphics[scale=0.08]{ORCIDiD.png}}, 
Scott M. Ransom$^{20}$\href{https://orcid.org/0000-0001-5799-9714}{\includegraphics[scale=0.08]{ORCIDiD.png}},
Qitong Ruan$^{17}$,
Ingrid Stairs$^{21}$\href{https://orcid.org/0000-0001-9784-8670}{\includegraphics[scale=0.08]{ORCIDiD.png}},
Chao-Wei Tsai$^{1}$\href{https://orcid.org/0000-0002-9390-9672}{\includegraphics[scale=0.08]{ORCIDiD.png}},
Pei Wang$^{1}$\href{https://orcid.org/0000-0002-3386-7159}{\includegraphics[scale=0.08]{ORCIDiD.png}},
Long Wang$^{22,23}$\href{https://orcid.org/0000-0001-8713-0366}{\includegraphics[scale=0.08]{ORCIDiD.png}},
Na Wang$^{24}$\href{https://orcid.org/0000-0002-9786-8548}{\includegraphics[scale=0.08]{ORCIDiD.png}},
Qingdong Wu$^{21}$,
Jianping Yuan$^{24}$\href{https://orcid.org/0000-0002-5381-6498}{\includegraphics[scale=0.08]{ORCIDiD.png}},
Jie Zhang$^{25}$,
Weiwei Zhu$^{1}$\href{https://orcid.org/0000-0001-5105-4058}{\includegraphics[scale=0.08]{ORCIDiD.png}},
Yongkun Zhang$^{1}$\href{https://orcid.org/0000-0002-8744-3546}{\includegraphics[scale=0.08]{ORCIDiD.png}}
and Di Li$^{1*}$\href{https://orcid.org/0000-0003-3010-7661}{\includegraphics[scale=0.08]{ORCIDiD.png}}
}

\affiliation{
 $^{1}$ {National Astronomical Observatories, Chinese Academy of Sciences, A20 Datun Road, Chaoyang District, Beijing 100101, China}\\
\textcolor{blue}{leizhang996@nao.cas.cn; dili@nao.cas.cn}\\
$^{2}$ {Centre for Astrophysics and Supercomputing, Swinburne University of Technology, P.O. Box 218, Hawthorn, VIC 3122, Australia}\\
$^{3}$ {Max-Planck-Institut f{\"u}r Radioastronomie, Auf dem H{\"u}gel 69, 53121 Bonn, Germany}\\
$^{4}$ {INAF – Osservatorio Astronomico di Cagliari, Via della Scienza 5, I-09047 Selargius (CA), Italy}\\
$^{5}$ Guizhou Radio Astronomical Observatory, Guizhou University Guiyang 550025, People's Republic of China\label{aff:GZRO}\\
$^{6}$ College of Astronomy and Space Sciences, University of Chinese Academy of Sciences, Chinese Academy of Sciences, Beijing 100101, P.~R.~China\label{aff:UCAS}\\
$^{7}$ Key Laboratory of Radio Astronomy and Technology, Chinese Academy of Sciences, Beijing 100101, People’s Republic of China\label{aff:keylab}\\
$^{8}$ {Department of Physics, CCIS 4-183, University of Alberta, Edmonton, AB, T6G 2E1, Canada}\\
$^{9}$ {Dipartimento di Fisica e Astronomia ``Augusto Righi'', Alma Mater Studiorum Universit\`a di Bologna, via Piero Gobetti 93/2, I-40129 Bologna, Italy}\\
$^{10}$ {INAF-Osservatorio di Astrofisica e Scienze dello Spazio di Bologna, Via Piero Gobetti 93/3 I-40129 Bologna, Italy}\\
$^{11}$ {Yunnan Observatories, Chinese Academy of Sciences, 396 Yangfangwang, Guandu District, Kunming 650216, People’s Republic of China}\\
$^{12}$ {Key Laboratory for the Structure and Evolution of Celestial Objects, Chinese Academy of Sciences, 396 Yangfangwang, Guandu District, Kunming 650216, China}\\
$^{13}$ {Center for Astronomical Mega-Science, Chinese Academy of Sciences, 20A Datun Road, Chaoyang District, Beijing 100012, China}\\
$^{14}$ {Purple Mountain Observatory, Chinese Academy of Sciences, Nanjing 210023, China}\\
$^{15}$ {School of Science, Western Sydney University, Locked Bag 1797, Penrith, NSW 2751, Australia}\\
$^{16}$ {National Supercomputing Center in Jinan, Qilu University of Technology, 28666 East Jingshi Road, Licheng District, Jinan 250103, China}\\ 
$^{17}$ {Jinan Institute of Supercomputing Technology, 28666 East Jingshi Road, Licheng District, Jinan 250103, China}\\
$^{18}$ ASTRON, the Netherlands Institute for Radio Astronomy, Oude Hoogeveensedijk 4, 7991 PD Dwingeloo, The Netherlands\\
$^{19}$ Anton Pannekoek Institute for Astronomy, University of Amsterdam, Science Park 904, 1098 XH, Amsterdam, The Netherlands\\
$^{20}$ NRAO, 520 Edgemont Road, Charlottesville, VA 22903, USA\\
$^{21}$ Dept. of Physics and Astronomy, University of British Columbia, 6224 Agricultural Road, Vancouver, BC V6T 1Z1, Canada.\\
$^{22}$ School of Physics and Astronomy, Sun Yat-sen University, Daxue Road, Zhuhai, 519082, China \\
$^{23}$ CSST Science Center for the Guangdong-Hong Kong-Macau Greater Bay Area, Zhuhai, 519082, China\\
$^{24}$ {Xinjiang Astronomical Observatory, Chinese Academy of Sciences, 150 Science 1-Street, Urumqi, Xinjiang 830011, China}\\
$^{25}$ College of Physics and Electronic Engineering, Qilu Normal University, 2 Wenbo Road, Zhangqiu District, Jinan 250200, China\\
}

\begin{abstract}
We report on a comprehensive multi-wavelength study of the pulsars in the globular cluster (GC) M5, including the discovery of M5G, a new compact non-eclipsing ``black widow'' pulsar. Thanks to the analysis of 34 years of radio data taken with the FAST and Arecibo telescopes, we obtained new phase-connected timing solutions for four pulsars and improved those of the other three.
These have resulted in, among other things: a) much improved proper motions for five pulsars, with transverse velocities (relative to the cluster) that are smaller than their respective escape velocities; b) 3-$\sigma$ and 1.5-$\sigma$ detections of Shapiro delays in M5F and M5D, respectively; c) greatly improved measurement of the periastron advance in M5B, whose value of $\dot{\omega} = 0.01361(6)^\circ$ implies that M5B is still likely to be a heavy ($m_{\rm p} = 1.981^{+0.038}_{-0.088} \, \rm M_{\odot}$) neutron star.
The binary pulsars M5D, E and F are confirmed to be in low-eccentricity binary systems, the low-mass companions of which are newly identified to be He white dwarfs using Hubble Space Telescope data. Four pulsars are also found to be associated with X-ray sources. Similarly to the eclipsing pulsar M5C, M5G shows little or no non-thermal X-ray emission, indicative of weak synchrotron radiation produced by intra-binary shocks. All the seven pulsars known in M5 have short spin periods ($<$ 8\,ms) and five are in binary systems with low orbital eccentricities. These characteristics differ from the overall GC pulsar population, but confirm the expectations for the pulsar population in a cluster with a small rate of stellar encounters per binary system.
\end{abstract}


\section{Introduction}\label{sec:introduction}
Globular clusters (GCs) are among the most prolific targets for radio pulsar searches. Since the discovery of the first GC pulsar (PSR B1821$-$24A in M28; \citealt{Lyne+1987}), 292 pulsars have been found in 38 different clusters\footnote{\url{https://www3.mpifr-bonn.mpg.de/staff/pfreire/GCpsr.html}} and these numbers are rapidly growing, especially in the last few years. 
More than 80\% of the GC pulsars are millisecond pulsars (MSPs, defined as $P < 30$\,ms), in stark contrast with the 10\% MSP fraction of the general Galactic population\footnote{\url{http://astro.phys.wvu.edu/GalacticMSPs/GalacticMSPs.txt}}. 
Such a contrast can be attributed to the peculiar conditions affecting stellar evolution in GCs, particularly the large age of the stellar population and the high stellar number densities ($10^{3}$--$10^{7}$~pc$^{-3}$) in GC cores, which are many orders of magnitude higher than that in the Galactic field, except for the Galactic centre. The resulting high stellar interaction rate favors the dynamical formation of low-mass X-ray binary (LMXB) systems \citep{Clark75}.

In such interactions, old isolated neutron stars (NS) in the GC, which had become undetectable after crossing the pulsar ``death line'' in the $P$--$\dot{P}$ diagram during their spin-down, can become bound to a low-mass star in a close orbit, and thus become part of an LMXB system. This can happen via tidal capture \citep{Fabian75}, by exchange encounters (in which the lighter component of a binary system gets replaced by another, most often heavier, star; \citealt{Hills76}), or by other mechanisms \citep{Sutantyo75}. These LMXB formation mechanisms probably happen at a rate that is proportional to the stellar encounter rate $\Gamma$ of the cluster \citep{Verbunt87}. 

In an LMXB, the NS can be re-activated by being spun-up (or ``recycled'') to rotation periods of just a few milliseconds through a Gyr-long phase of mass accretion from a light, unevolved donor star, a process that also circularizes the orbit. When accretion stops, we have a fast-spinning radio MSP binary with a low-mass companion in a nearly circular orbit \citep[e.g.,][]{Bhattacharya+1991}, as observed for the population of binary MSPs in the Galactic disk. The evolution of LMXBs in GCs is generally identical. A prime example is 47 Tucanae, where all the 29 pulsars currently known are MSPs with periods smaller than 8\,ms, and 17 of them are in binary systems with the expected characteristics of the recycling model \citep[e.g.][]{Ridolfi+2016,Freire+2017}.
 
 However, in some GCs with a large interaction rate per binary $\gamma$, these already recycled binaries can be disturbed through further strong stellar encounters. This can lead to unbinding of the systems (thus producing a larger fraction of isolated MSPs) or, if the disruption happens during the LMXB stage, to partially recycled pulsars with higher B-fields \citep{Verbunt_Freire2014}.
 In some cases, an exchange encounter may replace the remnant of the star that recycled the pulsar with a heavier degenerate star,  which results in a highly eccentric system composed of a pulsar and a heavy companion,
 such as PSR~B2127+11C in M15 \citep{Jacoby+2006}, PSR~J1835$-$3259A in NGC 6652 \citep{DeCesar+2015} and PSR~J0514$-$4002E in NGC 1851 \citep{Ridolfi+2022}.
 The high-$\gamma$ GCs containing these exotic binaries are high-density clusters, most of which are designated as core-collapsed. Their pulsar populations are remarkably different from GCs like 47~Tuc in their binary parameters.

\begin{figure*}
    \centering
    \includegraphics[width=0.9\textwidth]{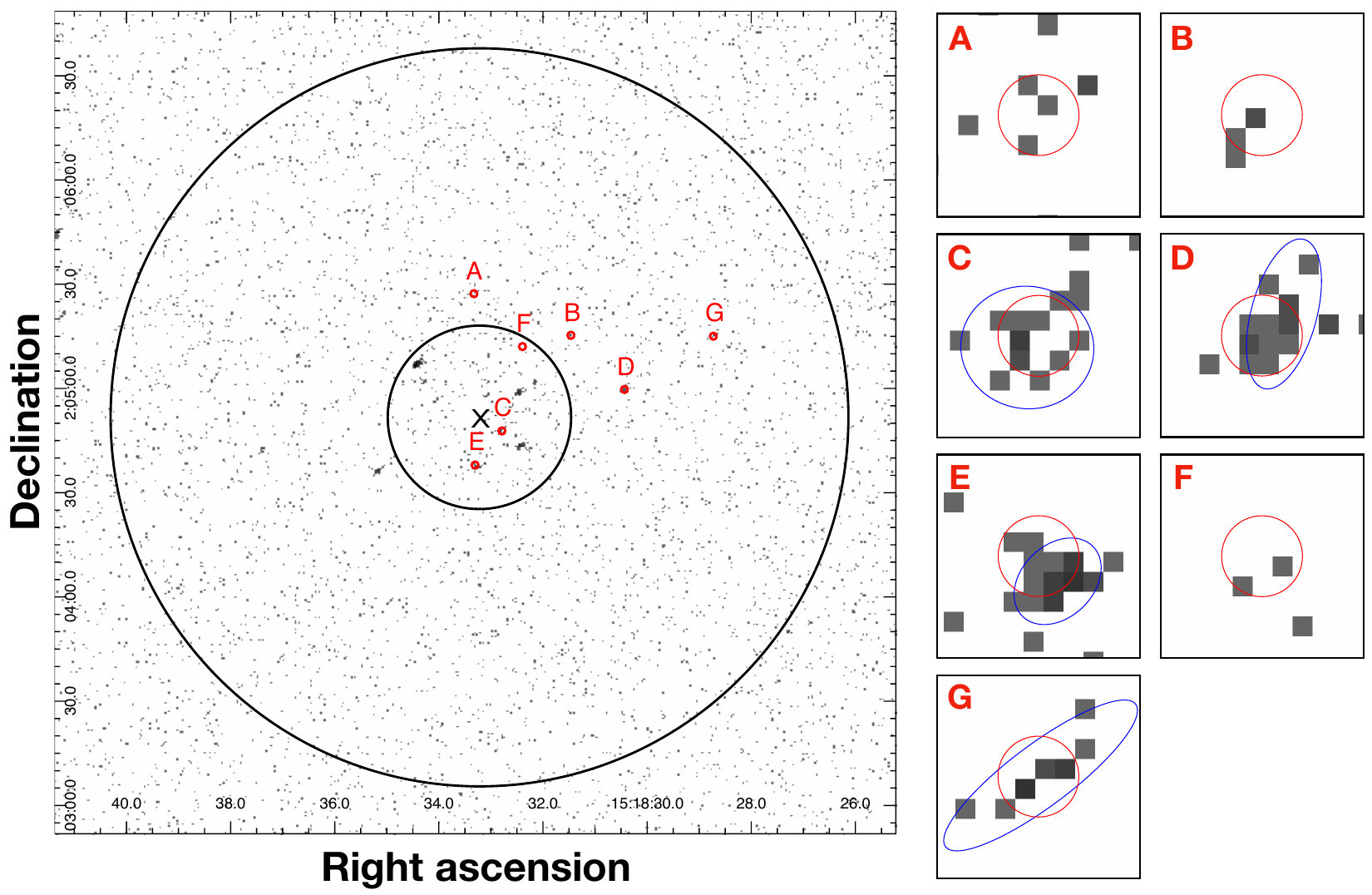}
    \caption{{\it Left}: Chandra X-ray image of M5 in the 0.3--8 keV energy band. The two concentric black circles with radii of $0\farcm44$ and $1\farcm77$ show the core radius and half-light radius, respectively, centered at R.A.=15:18:33.22, Dec=+02:04:51.7 \citep[black cross;][]{Goldsbury+2010}. The MSP locations are labeled with 1-arcsec-radius red circles centered at their corresponding timing positions, with pulsar names annotated above. {\it Right}: zoom-in images of M5 MSPs in 5\arcsec$\times$5\arcsec\ boxes. Blue ellipses show the source detections by {\tt wavdetect} (Table~\ref{tab:x-ray_cp}). X-ray counterparts are found to MSPs C, D, E, and G, whereas no X-ray sources are detected around MSPs A, B, and F.}
    \label{fig:chandra_m5}
\end{figure*}

A GC with a very small $\gamma$ is M5 (NGC~5904), a bright  ($V\approx5.6$) GC at a distance $d$ of 7.5\,kpc~\citep{Harris+2010}, the nominal center of which is at equatorial coordinates of $\alpha_{\rm J2000}=$ 15$\h$\,18$\m$\,33.214$\s$,  $\delta_{\rm J2000}=+02\degr\,04\arcmin\,51.80\arcsec$ \citep{Miocchi+2013} in the northern constellation Serpens. The GC core and half-light radii are $\theta_c = 0.44$ and $\theta_h = 1.77$ arcmin, respectively. As it was visible from the 305-m Arecibo radio telescope, in Puerto Rico (USA), it has been extensively observed with that telescope since 1989 at central frequencies of 430 MHz and 1400 MHz. These observations have led to the discovery of five pulsars \citep{Anderson+1997,Hessels+2007,2003AAS...203.5307M}. 

Its declination makes M5 a promising target for pulsar searches with the Five-hundred-meter Aperture Spherical radio Telescope (FAST, \citealt{Nan11IJMPD, Li18IMMag}), which can observe  over 40 GCs with declination between $-14^{\circ}$ and $ 66^{\circ}$ \citep{Zhang+2016}. With FAST, \cite{Pan+2021} discovered the sixth pulsar in M5 (M5F), which also has a spin period shorter than 8 ms and a circular orbit, as expected for a low-$\gamma$ cluster. 

Except for M5A, the other five (M5B, C, D, E and F) known pulsars in M5 spin very fast and are members of binary systems with low eccentricities and low-mass companions, as might be expected from the low $\gamma$ of M5. Their dispersion measures (DMs) are between 29.4 to 30.1\,pc cm$^{-3}$ and their 1400\,MHz flux densities are estimated to be in the range 0.01--0.12\,mJy, although the observed brightness of the pulsars varies significantly due to diffractive scintillation. Only pulsars A, B, and C have published radio timing phase-connected solutions and optical observations \citep{Anderson+1997, Freire+2008, Pallanca+2014}.

In this paper, we present the discovery of a new pulsar (M5G) with FAST. We then present the timing solution of all seven pulsars, combining FAST and Arecibo data, spanning 34 years. The M5 pulsar population was analyzed with multi-wavelength observations from radio, optical, and X-ray instruments. In Section 2, we describe the observations and data reduction. The discovery, timing solutions, and multi-wavelength emission properties of the pulsars in M5 are presented in Section 3. In Section 4, we discuss the implications of our findings and summarize our conclusions in Section 5.

\section{Observations} 
The radio data set is composed of new FAST observations and archival Arecibo data for the GC. A summary of all the radio observations used in this paper is listed in Table~\ref{tab:M5ephem}. We also utilized archival high-resolution near-UV and optical images from the {\it Hubble Space Telescope (HST)} and high-energy data from the {\it Chandra} X-ray Observatory. In the following section, we describe the observations made with each telescope.

\subsection{FAST Observations} 
We have carried out 33 observations of M5 between 2020 November 16 and 2022 December 14 using the central beam (the beam width is 3\arcmin\ at 1250\,MHz) of the FAST 19-beam receiver through three observing projects (PT2020\_0074, PT2021\_0061, PT2022\_0062). For the project PT2020\_0074, the telescope was pointed at  M5B, at $15^{\rm h}\;18^{\rm m}\;31^{\rm s}.46$, $+02^{\circ}\;05^{'}\;15^{''}.30$~\citep{Freire+2008}. For the latter two projects, the telescope was pointed at the nominal cluster center: $15^{\rm h}\;18^{\rm m}\;33^{\rm s}.22$, $+02^{\circ}\;04^{'}\;51^{''}.7$~\citep{Goldsbury+2010}. All the FAST observations were recorded with 8-bit sampling every 49\,${\mu}s$ in pulsar search mode. In all the observations, the observing band from 1000\,MHz to 1500\,MHz was split into 4096 frequency channels and due to bandpass roll-off the effective band is from 1050\,MHz to 1450\,MHz. 

\subsection{Arecibo Observations}
The GC M5 was first observed by Arecibo with the 430-MHz Carriage House line feed receiver in the years between 1989--1994 \citep{Anderson+1997}. From this dataset, we were able to retrieve times-of-arrival (ToAs) only for the pulsars M5A and B. 
Arecibo observations of M5 resumed in 2001, initially using the ``old'' Gregorian L-band receiver (2001-2003), and later using the ``new'', and more sensitive Gregorian L-band receiver (2003-2008). For details on the exact observing set-up and recording modes, we refer to \cite{Freire+2008}. From the latter two datasets, we were able to retrieve most of the observations, archived as \textsc{presto}\footnote{\url{https://github.com/scottransom/presto}} sub-banded files, i.e. search-mode files that were greatly reduced in size by summing groups of frequency channels after de-dispersing at the DMs of pulsars A-E. 

\subsection{Hubble Space Telescope}
We analyzed deep and high-resolution near-UV and optical observations obtained with the Advanced Camera for Surveys and with the Wide Field Camera 3 onboard the Hubble Space Telescope through six different filters: F275W, F336W, F390W, F435W, F606W, F814W. These observations were carried out as part of different HST proposals:
GO-12517 (PI: Ferraro), 
GO-11615 (PI: Ferraro), 
GO-15875 (PI: Bellini), 
GO-10615 (PI: Anderson), 
and GO-13297 (PI: Piotto). 

The data reduction was performed using  DAOPHOT IV \citep{1987PASP...99..191S, 1994PASP..106..250S} on the calibrated images and adopting the so-called ``UV-route". Details on the data reduction routines can be found in \citet[see also \citealt{2023ApJ...948...84C,2020ApJ...905...63C,2020ApJ...895...54C}]{2021NatAs...5.1170C}. The instrumental magnitudes were calibrated to the VEGAMAG photometric system by using appropriate zero points and aperture corrections. We transformed the instrumental positions of the sources to the International Celestial Reference System (ICRS) by cross-correlating our catalog with the Gaia Data Release 3 catalogue of stars \citep{2022arXiv220800211G}. The residuals of this transformation returned a combined r.m.s. of $\sim 15$ mas, which was adopted as the $1\sigma$ astrometric accuracy. 

\subsection{Chandra X-ray observatory}
M5 has been observed by the {\it Chandra X-ray Observatory} (CXO) with an exposure time of 44.7 ks in the FAINT data mode (Observation ID 2676). Figure~\ref{fig:chandra_m5} (left panel) shows the X-ray image of M5 in the band 0.3-8\,keV. The X-ray counterpart to M5C was identified by \citet{Zhao+2022}, whereas no X-ray counterparts were found for M5A and B. 
Here, using our new timing positions of M5 MSPs, particularly the four MSPs with newly derived timing solutions, we searched for their X-ray counterparts using the Chandra observation\footnote{\textcolor{red}{The X-ray analysis in this paper employs a Chandra dataset, obtained by the Chandra X-ray Observatory, contained in~\dataset[Chandra Data Collection (CDC) 160]{\url{https://doi.org/10.25574/cdc.160}}}}.

\section{ Analyses and Results}
\begin{table*}[]
\centering
\scriptsize
\caption{Parameters for M5 pulsars}
\setlength{\tabcolsep}{0.6mm}{
\begin{tabular}{lcccc}
\hline
Pulsar name                                                               & M5A                      & M5B                    & M5C                      & M5D                     \\\hline
                                                                          &                          &                        &                          &                         \\
\textbf{Observation and data reduction parameters}                                                   &                        &                          &                          &                         \\
Solar System Ephemeris\dotfill                                                    & DE440                    & DE440                  & DE440                    & DE440                   \\
Timescale\dotfill                                                                 & TDB                      & TDB                    & TDB                      & TDB                     \\
Binary model\dotfill                                                              & --                       & DDGR                   & BTX                      & ELL1                    \\
Reference epoch for $\alpha$, $\delta$ and $\nu$ (MJD)\dotfill                    & 54000                    & 54000                  & 54000                    & 59500                   \\
MJD range of 430 MHz Arecibo\dotfill                                              & 47635--49354             & 47635--49394           & --                       & --                      \\
Data span of 430 Arecibo (days)\dotfill                                           & 1720                     & 1760                   & --                       & --                      \\
MJD range of L-band Arecibo\dotfill                                               & 52087--54685             & 52087--55814           & 52483--59831             & 53058--55814            \\
Data span of L-band Arecibo (days)\dotfill                                        & 2599                     & 3728                   & 7349                     & 2757                    \\
MJD range of L-band FAST\dotfill                                                  & 59169--59927             & 59169--59927           & 59169--59927             & 59169--59927            \\
Data span of L-band FAST (days)\dotfill                                           & 759                      & 759                    & 759                      & 759                     \\
Number of TOAs at 430 MHz Arecibo\dotfill                                         & 86                       & 81                     & --                       & --                      \\
Residual RMS at  430 MHz  Arecibo ($\mu s$)\dotfill                               & 48.133                   & 112.715                & --                       & --                      \\
Number of TOAs at L-band Arecibo\dotfill                                          & 1340                     & 222                    & 1394                     & 97                      \\
Residual RMS at  L-band Arecibo ($\mu s$)\dotfill                                 & 8.504                    & 71.272                 & 11.224                   & 23.286                  \\
Number of TOAs at L-band FAST\dotfill                                             & 132                      & 125                    & 309                      & 165                     \\
Residual RMS at  L-band FAST ($\mu s$)\dotfill                                    & 1.096                    & 22.092                 & 1.664                    & 12.118                  \\
Reduced $\chi^{2}$\dotfill                                                  & 1.011                    & 1.023                  & 1.008                    & 1.015                   \\
\\
\textbf{Measured Parameters}                                              &                          &                        &                          &                         \\
Flux density at 1.25 GHz (mJy)$^a$\dotfill                                        & 0.137(8)                 & 0.033(4)               & 0.044(2)                 & 0.017(1)                \\
Pulse width at 10\% of peak, $W_{10}$ (ms)$^a$\dotfill                            & 0.764                    & 3.715                  & 1.303                    & 1.851                   \\
Pulse width at 50\% of peak, $W_{50}$ (ms)$^a$\dotfill                            & 0.367                    & 1.024                  & 1.077                    & 0.557                   \\
Right Ascension, $\alpha$ (J2000)\dotfill                                         & 15:18:33.32293(1)        & 15:18:31.4629(2)       & 15:18:32.78978(2)        & 15:18:30.43512(6)       \\
Declination, $\delta$ (J2000)\dotfill                                             & +02:05:27.4314(4)        & +02:05:15.306(7)       & +02:04:47.7850(6)        & +02:04:59.712(4)        \\
Proper motion in $\alpha$, $\mu_{\alpha}$ (mas yr$^{-1}$, J2000)\dotfill          & 4.13(1)                  & 4.0(2)                 & 4.25(2)                  & 4.3(2)                  \\
Proper motion in $\delta$, $\mu_{\delta}$ (mas yr$^{-1}$, J2000)\dotfill          & $-$10.03(3)              & $-$10.9(6)             & $-$9.78(5)               & $-$10.2(6)              \\
Spin frequency, $\nu$ (s$^{-1}$)\dotfill                                          & 180.063624055099(2)      & 125.8345875794094(2)   & 402.588227988547(4)      & 334.67467287054(2)      \\
Spin frequency derivative, $\dot{\nu}$ (s$^{-2}$)\dotfill                         & $-$1.33902(3)$\times10^{-15}$ & 5.238(3)$\times10^{-17}$ & $-$4.22684(4)$\times10^{-15}$ & -2.48423(6)$\times10^{-15}$ \\
Second time derivative of spin frequency, $\ddot{\nu}$ ($\text{s}^{-3}$)\dotfill & $-$3.6(2)$\times10^{-27}$      & --                  & $-$1.3(1)$\times10^{-26}$         & --             \\
Dispersion Measure, DM (cm$^{-3}$ pc)$^a$\dotfill                                 & 30.0546(5)               & 29.469(4)              & 29.3109(3)               & 29.371(2)               \\
Orbital Period, $P_{b}$ (days)\dotfill                                            & --                       & 6.85845358(4)          & 0.08682882871(7)         & 1.2220885130(2)         \\
Projected semi-major axis, $\chi$ (lt-s)\dotfill                                  & --                       & 3.048266(3)            & 0.0573202(1)             & 1.596065(4)             \\
Epoch of periastron, T$_{0}$ (MJD)\dotfill                                      & --                       & 54004.02035(3)         & 52850.0043442(2)         & --        \\
Epoch of the ascending node, T$_{\rm asc}$ (MJD)\dotfill                              & --                       & --                     & --         & 59189.3836465(2)        \\
$e \sin \omega,  \varepsilon_{1} (10^{-5})$\dotfill                                & --                       & --                    & --                       & 0.25(30)                \\
$e \cos \omega,  \varepsilon_{2} (10^{-5})$\dotfill                                & --                       & --                    & --                    & $-$0.14(19)               \\
Orbital eccentricity, $e$\dotfill                                                 & --                       & 0.137839(2)            & 0.00$^d$                       & --                      \\
Longitude of periastron, $\omega$ (deg)\dotfill                                   & --                       & 359.894(2)             & 0.00$^d$                       & --                      \\
Rate of advance of periastron, $\dot{\omega}$ (deg yr$^{-1}$)\dotfill               & --                     & 0.01361(6)             & --                       & --                      \\
Time derivative of $\chi$, $\dot{\chi}$ (10$^{-14}$ lt-s s$^{-1}$)\dotfill        & --                       & 2.16(3)                & --                       & --                      \\
Time derivative of $P_{b}$, $\dot{P}_{b}$ (10$^{-12}\text{s}~\text{s}^{-1}$)\dotfill & --                    & $-$1.7(18)             & --                       & --                      \\
Orbital frequency, $f_{b} (\text{s}^{-1})$\dotfill                                 & --                      & --          &$1.3329759535(7)\times10^{-4}$       & --                       \\
1st orbital frequency derivative, $f_{b}^{1} (\text{s}^{-2})$\dotfill              & --                      & --          &$-6.3(5)\times10^{-20}$              & --                       \\
2st orbital frequency derivative, $f_{b}^{2} (\text{s}^{-3})$\dotfill              & --                      & --          &$3.6(2)\times10^{-27}$               & --                       \\
3st orbital frequency derivative, $f_{b}^{3} (\text{s}^{-4})$\dotfill              & --                      & --          &$-7.8(5)\times10^{-35}$              & --                       \\
4st orbital frequency derivative, $f_{b}^{4} (\text{s}^{-5})$\dotfill              & --                      & --          &$9.2(7)\times10^{-43}$               & --                       \\
5st orbital frequency derivative, $f_{b}^{5} (\text{s}^{-6})$\dotfill              & --                      & --          &$-6.2(5)\times10^{-51}$              & --                       \\
6st orbital frequency derivative, $f_{b}^{6} (\text{s}^{-7})$\dotfill              & --                      & --          &$1.8(2)\times10^{-59}$               & --                       \\
\\

\textbf{Derived Parameters}                                               &                          &                        &                          &                         \\
Position perpendicular offset from center (pc)$^b$\dotfill                            & 1.300                    & 1.286                  & 0.274                    & 1.546                   \\
Spin period, $P$ (ms)\dotfill                                                     & 5.55359254401101(5)      & 7.94694065627177(2)    & 2.48392757283615(2)      & 2.9879763276464(2)      \\
Spin period first time derivative, $\dot{P}~(10^{-20}\text{s}~\text{s}^{-1})$\dotfill  & 4.12985(9)          & $-$0.3308(2)             & 2.60791(3)               & 2.60791(3)              \\
Mass function, $f (\rm M_{\odot})$\dotfill                                            & --                       & 0.000646530            & 0.0000268213             & 0.00292300           \\
Minimum companion mass, $M_{\rm c,min} (\rm M_{\odot})^c$\dotfill                           & --                       & 0.1114                 & 0.03723                  & 0.1907                  \\
Median companion mass, $M_{\rm c,med} (\rm M_{\odot})^c$\dotfill                            & --                       & 0.1297                 & 0.0431                   & 0.2233                  \\ \hline
\end{tabular}}
\label{tab:M5ephem}
\end{table*}

\begin{table*}
\label{tab:parTab}
\centering
\scriptsize
\setlength{\tabcolsep}{5mm}{
\begin{tabular}{lccc}
\hline
Pulsar name                                                        & M5E                    & M5F                      & M5G                     \\\hline
                                                                   &                        &                          &                         \\
\textbf{Observation and data reduction parameters}                 &                        &                          &                         \\
Solar System Ephemeris\dotfill                                            & DE440                  & DE440                    & DE440                    \\
Timescale\dotfill                                                         & TDB                    & TDB                      & TDB                     \\
Binary model\dotfill                                                      & ELL1                   & ELL1                     & ELL1                     \\
Reference epoch for $\alpha$, $\delta$ and $\nu$ (MJD)\dotfill            & 59500                  & 59500                    & 59500                    \\
MJD range of 430 MHz Arecibo\dotfill                                      & --                     & --                       & --                      \\
Data span of 430 Arecibo (days)\dotfill                                   & --                     & --                       & --                      \\
MJD range of L-band Arecibo\dotfill                                       & 53029-54685            & --                       & --                      \\
Data span of L-band Arecibo (days)\dotfill                                & 1656                   & --                       & --                      \\
MJD range of L-band FAST\dotfill                                          & 59169--59927           & 59169--59927             & 59169--59927            \\
Data span of L-band FAST (days)\dotfill                                   & 759                    & 759                      & 759                     \\
Number of TOAs at 430 MHz Arecibo\dotfill                                 & --                     & --                       & --                         \\
Residual RMS at  430 MHz  Arecibo ($\mu s$)\dotfill                       & --                     & --                       & --                         \\
Number of TOAs at L-band Arecibo\dotfill                                  & 104                    & --                       & --                         \\
Residual RMS at  L-band Arecibo ($\mu s$)\dotfill                         & 10.141                 & --                       & --                         \\
Number of TOAs at L-band FAST\dotfill                                     & 188                    & 110                      & 161                      \\
Residual RMS at  L-band FAST ($\mu s$)\dotfill                            & 3.431                  & 6.337                    & 6.747                    \\
Reduced $\chi^{2}$\dotfill                                          & 1.003                  & 1.004                   & 1.013                   \\

\\
\textbf{Measured Parameters}                                       &                        &                          &                          \\
Flux density at 1.25 GHz (mJy)$^a$\dotfill                                 & 0.022(3)               & 0.012(3)                 & 0.008(1)                 \\
Pulse width at 10\% of peak, $W_{10}$ (ms)$^a$\dotfill                     & 2.288                  & 2.102                    & 0.555                    \\
Pulse width at 50\% of peak, $W_{50}$ (ms)$^a$\dotfill                     & 0.875                  & 1.711                    & 0.133                    \\
Right Ascension, $\alpha$ (J2000)\dotfill                                  & 15:18:33.3027(4)       & 15:18:32.3948(3)         & 15:18:28.72895(9)        \\
Declination, $\delta$ (J2000)\dotfill                                      & +02:04:37.954(2)       & +02:05:12.07(1)         & +02:05:15.040(4)         \\
Proper motion in $\alpha$, $\mu_{\alpha}$ (mas yr$^{-1}$, J2000)\dotfill   & 4.04(12)               & --                      & --                       \\
Proper motion in $\delta$, $\mu_{\delta}$ (mas yr$^{-1}$, J2000)\dotfill   & $-$10.35(26)             & --                      & --                       \\
Spin frequency, $\nu$ (s$^{-1}$)\dotfill                                   & 314.23809500370(1)     & 376.7625236987(2)       & 363.61166596073(2)       \\
Spin frequency derivative, $\dot{\nu}$ (s$^{-2}$)\dotfill                  & $-$1.77140(4)$\times10^{-15}$ & $-$3.14(2)$\times10^{-15}$ & $-$1.649(3)$\times10^{-15}$ \\
Second time derivative of spin frequency, $\ddot{\nu}$ ($\text{s}^{-3}$)\dotfill & --                    & 5.2(15)$\times10^{-24}$  & --                  \\
Dispersion Measure, DM (cm$^{-3}$ pc)$^a$\dotfill                          & 29.310(1)              & 29.409(1)                & 29.3945(7)               \\
Orbital Period, $P_{b}$ (days)\dotfill                                     & 1.09693224035(8)       & 1.609518449(2)           & 0.1139283090(6)          \\
Projected semi-major axis, $\chi$ (lt-s)\dotfill                           & 1.1516157(8)           & 1.953013(2)              & 0.0359535(8)             \\
Epoch of the ascending node, T$_{\rm asc}$ (MJD)\dotfill                       & 59198.0057398(1)       & 59202.4806455(5)         & 59204.7170718(9)         \\
$e \sin \omega,  \varepsilon_{1} (10^{-5})$\dotfill                         & $-$0.78(15)            & $-$0.31(17)                & 0.00$^d$                     \\
$e \cos \omega,  \varepsilon_{2} (10^{-5})$\dotfill                         & 0.14(12)               & $-$0.25(16)                & 0.00$^d$                     \\
\\
\textbf{Derived Parameters}                                        &                        &                          &                          \\
Position perpendicular offset from center (pc)$^b$\dotfill                 & 0.502                  & 0.867                    & 2.591                    \\
Spin period, $P$ (ms)\dotfill                                              & 3.1823003509114(1)     & 2.654191797482(1)        & 2.7501867888583(1)       \\
Spin period first time derivative, $\dot{P}~(10^{-20}\text{s}~\text{s}^{-1})$\dotfill & 1.79390(4)  & 2.21(1)                  & 1.247(2)                 \\
Mass function, $f (\rm M_{\odot})$\dotfill                                       & 0.001362843          & 0.00308749               & 0.0000038445             \\
Minimum companion mass, $M_{\rm c,min}~(\rm M_{\odot})^c$\dotfill                    & 0.145                  & 0.1946                   & 0.01932                \\
Median companion mass, $M_{\rm c,med}~(\rm M_{\odot})^c$\dotfill                     & 0.1692                 & 0.2279                   & 0.0223                 \\ \hline
\multicolumn{4}{l}{{\bf Note:} Numbers in parentheses represent uncertainties on the last digit.}\\
\multicolumn{4}{l}{$^{a}$ Parameter derived from FAST data only.}\\ 
\multicolumn{4}{l}{$^{b}$ The pulsar’s position is offset from the nominal cluster center ($15^{\rm h}\;18^{\rm m}\;33^{\rm s}.22$, $+02^{\circ}\;04^{'}\;51^{''}.7$) for a cluster distance 7.5 kpc.}\\ 
\multicolumn{4}{l}{$^{c}$ The companion masses are derived from radio timing and assume a pulsar mass of 1.4$M_{\odot}$.}\\ 
\multicolumn{4}{l}{The minimum and median masses assume an inclination angle of $90^{\circ}$ and $60^{\circ}$, respectively.}\\
\multicolumn{4}{l}{$^{d}$ Because of the small semi-major axis and the presence of eclipses or dispersive delays, eccentricity measurements in black widow}\\
\multicolumn{4}{l}{systems have large uncertainties and are unreliable. In these systems, is expected that tidal dissipation keeps orbits circular. In this}\\
\multicolumn{4}{l}{work, we have assumed that this is indeed the case.}\\
\end{tabular}}
\end{table*}

\begin{figure*}[]
\centering 
\includegraphics[width=1.0\linewidth]{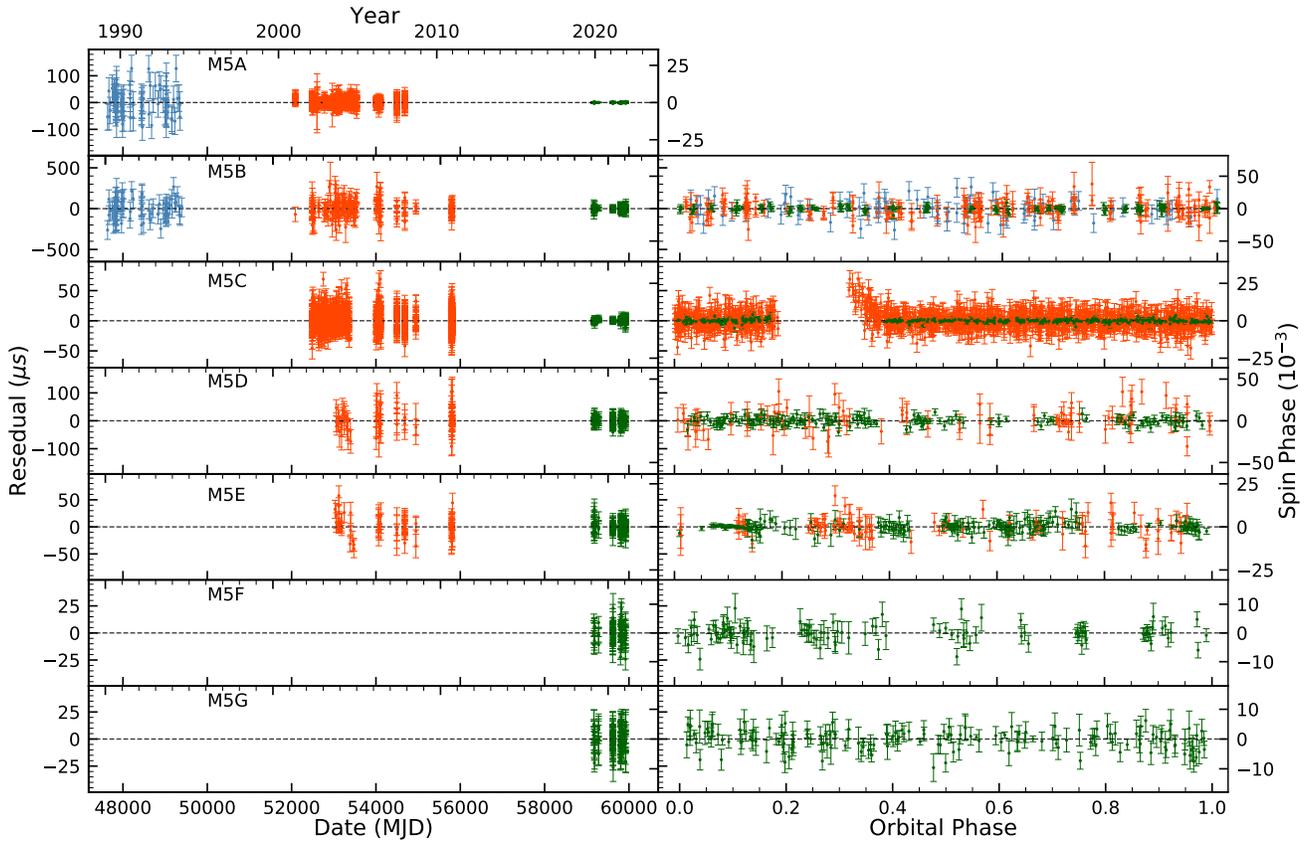}
\caption{Timing residuals of the seven MSPs in M5 obtained using the timing solutions in Table~\ref{tab:M5ephem}. The left panels show the post-fit timing residuals as a function of time, whereas the right panels show the post-fit timing residuals as a function of orbital phase for the binary pulsars (i.e., all except M5A). The blue, orange and green points represent observations taken with Arecibo at 430\,MHz, Arecibo at L-band and FAST at L-band, respectively.}
\label{fig.M5all_res}
\end{figure*}

\begin{figure*}[]
\centering 
\includegraphics[width=1.0\linewidth]{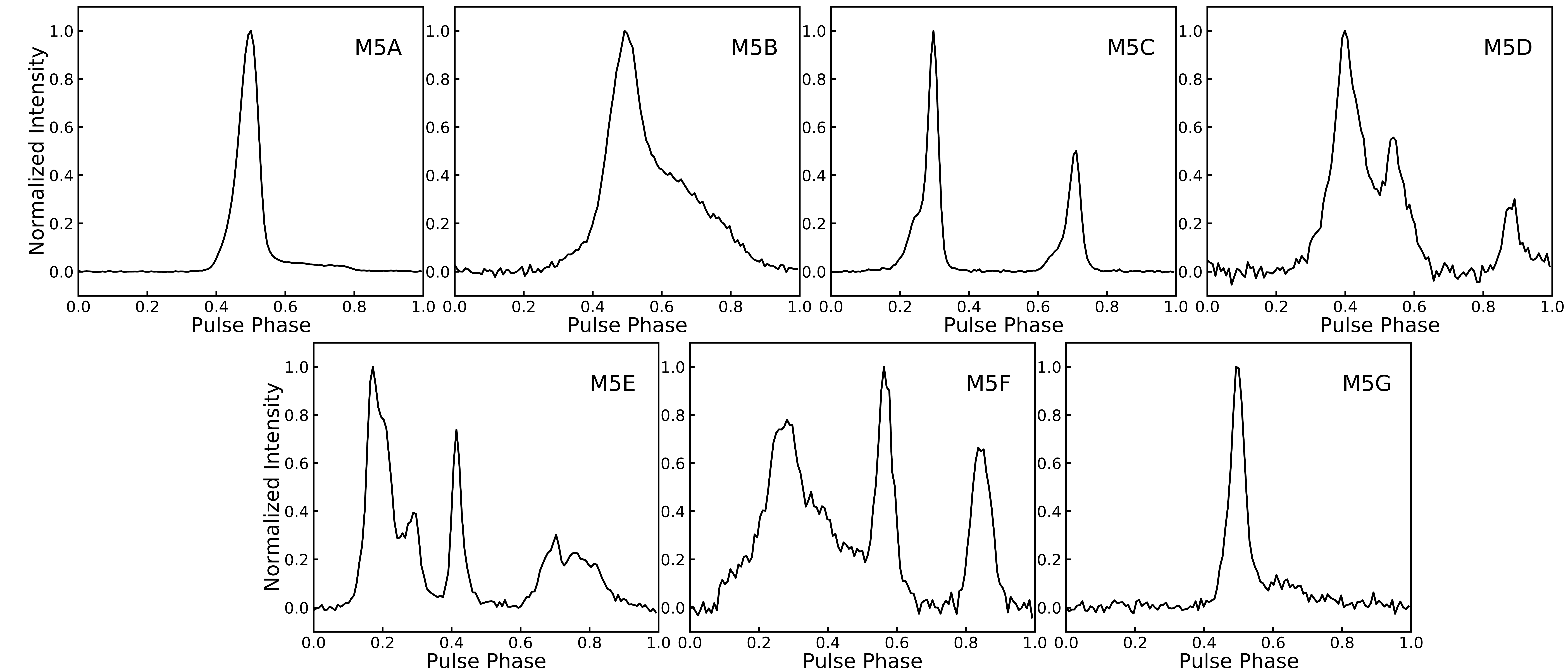}
\caption{Integrated profiles of the seven pulsars in M5 from FAST data at 1.25\,GHz. }
\label{fig.M5all_prof}
\end{figure*}

\subsection{Discovery of M5G and Timing Analysis}\label{subs_3.1}

We searched for new MSPs in the FAST observations using the \textsc{presto} software suite~\citep{Ransom+2002}, with a Fourier-domain jerk search with a maximum linear Fourier drift rate $z_{\rm max} = 200$ and a maximum Fourier jerk drift rate $w{\rm max} = 300$ (see ~\citealt{Andersen+2018}). We also split the long observations (more than two hours) into 15, 30 and 60 minute blocks so as to be sensitive to orbit with orbital periods as short as $\sim2$ hours)~\citep{Ng+2015}. We found a 2.7-ms pulsar signal at a DM of 29.4 \,pc cm$^{-3}$ in multiple observations, each time with a different associated acceleration, suggesting the presence of a binary motion. Being the seventh pulsar known in the cluster, the pulsar was named M5G (or PSR~J1518+0204G). Of the 33 follow-up observations, M5G was detected on 30 occasions, over the span of two years. The non-detections are most likely attributable to unfavorable scintillation.

In order to characterize M5G, we performed a radio timing analysis as follows. First, we folded each search-mode observation with 60-s sub-integrations and 128 profile bins using \textsc{dspsr}\footnote{\url{http://dspsr.sourceforge.net}}~\citep{vanStraten+2011}. We removed data affected by radio frequency interference (RFI) both in the frequency and time domains. After that, for each folded archive, we summed groups of frequency channels and sub-integrations so as to have sufficiently high S/N integrated profiles, which were then cross-correlated with a profile template to extract pulse times of arrival (ToAs) using the \textit{pat} routine
from the \textsc{psrchive}\footnote{\url{http://psrchive.sourceforge.net}\label{psrchive}}~\citep{Hotan+2004} package. The \textsc{tempo2}~\citep{Hobbs+2006} timing software was then used to develop a comprehensive timing model of the pulsar's behavior, including its position, rotation, and a binary model ELL1~\citep{Lange+2001} which assumes low orbital eccentricity. These procedures were iterated a few times until a phase-connected timing solution was obtained. The DM was measured with TEMPO2 using ToAs from multifrequency subbands. The timing solution of pulsar M5G is presented in Table~\ref{tab:M5ephem}. In Figure~\ref{fig.M5all_res}, the timing residuals are shown as functions of time and orbital phase: as can be seen, its radio pulsations can be detected at all orbital phases. M5G is a non-eclipsing ``black widow" type binary MSP, with an orbital period of 2.73 hours and with a companion mass of around 0.02\,$\rm M_{\odot}$, if we assume the pulsar has a mass of $\sim$ 1.4M$_{\odot}$.

\subsection{Timing of known MSPs}\label{subs_3.2}
Six pulsars had been discovered in previous observations of M5. The timing solutions of M5A, M5B and M5C were published by ~\cite{Freire+2008} and \cite{Pallanca+2014}, but have not been updated since. There are no published timing solutions for M5D, E and F, even though pulsars D and E were discovered in 2007. To time those pulsars, we first attempted to fold our data using the ephemerides from the ATNF pulsar catalog\footnote{\url{https://www.atnf.csiro.au/research/pulsar/psrcat/}}. However, we found that those ephemerides are not accurate enough to keep phase coherence in our data, except for M5A. Thus, to detect M5B to M5F for each epoch of observation, we performed a blind search for their periodic signal and then folded the data using the local ephemerides derived from the search. Once the detections were made, we then followed the procedures as described in section~\ref{subs_3.1} to obtain an accurate timing solution for all six pulsars. The timing solutions and residuals of these pulsars are summarized in Table~\ref{tab:M5ephem} and shown in Figure~\ref{fig.M5all_res}, respectively.  The timing parameters are reported in dynamical barycentric time (TDB), and we have used the Jet Propulsion Laboratory's DE 440 Solar System ephemeris \citep{Park2021} for all pulsars.

\subsection{Pulse profiles and Flux Densities}\label{subs_3.3}
To probe the pulsed emission properties of all seven pulsars, we again used programs from the {\sc psrchive} package. We constructed average pulse profiles by summing the FAST observations in time using the timing solution of the pulsar to ensure phase alignment. We present the average profiles at 1.25\,GHz in Figure~\ref{fig.M5all_prof}. The pulse widths at $50\%$ (W$_{50}$) and $10\%$ (W$_{10}$) of the pulse peak were measured using the program \textsc{pdv} from noise-free profiles obtained using the program \textsc{paas}, while we estimated each pulsar's flux density at 1.25\,GHz using the radiometer equation. The W$_{50}$, W$_{10}$, and flux densities estimates for the M5 pulsars are also listed in Table~\ref{tab:M5ephem}.

\subsection{Update on mass measurement for M5B}\label{subs_3.4}
One of the main results published by \cite{Freire+2008} was the measurement of the rate of periastron advance for M5B, $\dot{\omega} = 0.0142(7)^\circ \, \rm yr^{-1}$, which was made possible by its relatively large orbital eccentricity, $e = 0.138$, by far the largest among the known binary pulsars in the cluster. Assuming that this effect is purely relativistic (an assumption based on the non-detection of the companion to the pulsar in HST images, see also section~\ref{sec:optical} below), and that general relativity (GR) is the correct theory of gravity, this translated into a total mass $M_{\rm T} = 2.29 \pm 0.17 \, \rm M_{\odot}$. Given the low mass function of the system, a statistical analysis that assumes {\rm a priori} randomly aligned orbits implied a large mass for the pulsar, $M_{\rm T} = 2.08 \pm 0.19 \, \rm M_{\odot}$, with a 95\% probability of a mass above $1.72 \, \rm M_{\odot}$ and only a 0.77\% probability of a mass within the range of pulsar masses then known, $1.2$ to $1.44 \, \rm M_{\odot}$. This was an early indication that neutron stars could have masses substantially above $1.44 \, \rm M_{\odot}$. However, apart from their unlikelihood, there is no physical measurement that precludes the occurrence of a lower orbital inclination. In addition, in a large set of binary systems in globular clusters, some should be seen with a range of low orbital inclinations.

\begin{figure*}[]
\centering 
\includegraphics[width=.95\textwidth]{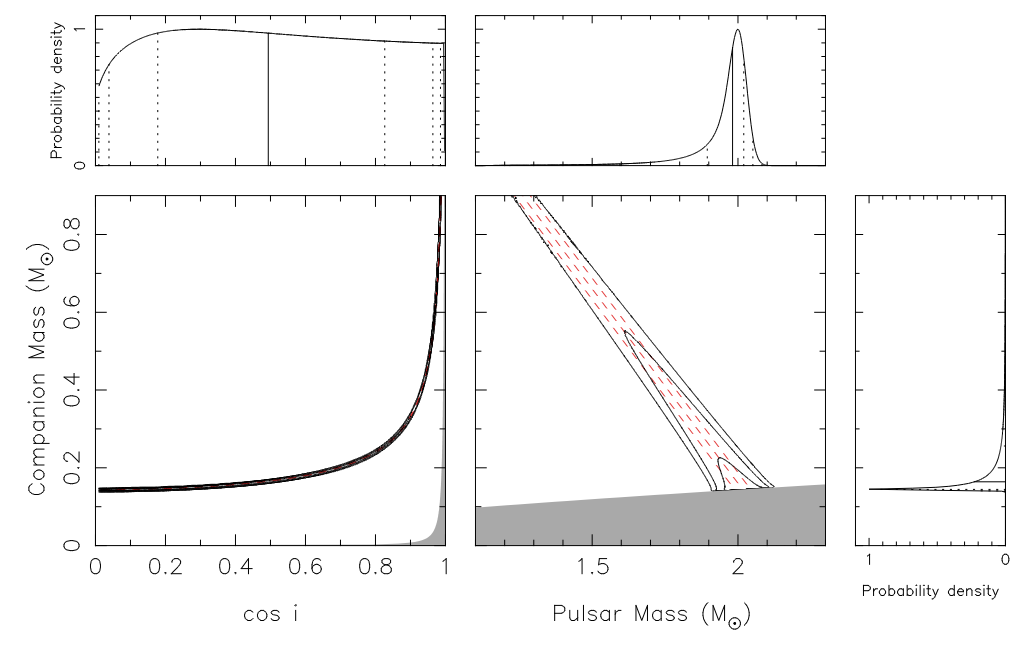}
\caption{Mass constraints for the M5B binary system. In the main panels, we show the 
 $m_{\rm c}$-$\cos i$ and $m_{\rm c}$-$m_{\rm p}$ planes. In the left panel, the grey zone is excluded because of $m_{\rm p}$ must be $>$ 0, in the right panel the grey zone is excluded because of $\sin i$ must be $\leq 1$. The contours include, respectively, 63.28\%, 95.45\% and 99.72\% of all probability in the 2-D posterior probability density functions (pdfs). The red dashed lines represent the median and $\pm 1$-$\sigma$ constraints on the total mass derived from the measurement of the rate of advance of periastron ($\dot{\omega}$). In the side panels, the curves show the 1-D pdfs for $\cos i$, $m_{\rm p}$ and (on the right panel) $m_{\rm c}$; these medians are indicated by the solid lines, the equivalent 1, 2 and 3-$\sigma$ percentiles are indicated by the dashed lines.}
\label{fig:mass_mass}
\end{figure*}

The inclusion of additional observations has greatly improved the astrometric, spin and orbital parameters of M5B. However, apart from an order-of-magnitude improvement of the rate of periastron advance ($\dot{\omega} = 0.01361(6)^\circ \, \rm yr^{-1}$, which is 1-$\sigma$ compatible with the value presented by \citealt{Freire+2008}), no additional relativistic effects have been detected. Such effects, like the Einstein delay ($\gamma_{\rm E}$) or the Shapiro delay \citep{Shapiro1964},  are required for determinations of the individual component masses.

In order to try and detect additional post-Keplerian parameters to allow estimates of the masses, we have made a $\chi^2$ map of the $m_{\rm c}$-$\cos i$ space, following the Bayesian procedure described in detail by \cite{Splaver+2002}. To sample the space, we used the DDGR timing solution \citep{DamourDeruelle1986}, which self-consistently accounts for all relativistic effects in the timing, even if they are present at a very low level, with the assumption that GR is the correct of gravity theory. Given the precise constraint on $\dot{\omega}$, we restrict the $m_{\rm c}$-$\cos i$ space being sampled to values of the total mass that are not too different from the new GR estimate of the total mass derived from $\dot{\omega}$.

The results for the 2-D posterior probability density functions (pdfs) in the $m_{\rm c}$-$\cos i$ and $m_{\rm c}$-$m_{\rm p}$ planes are depicted in the main panels of Fig.~\ref{fig:mass_mass}, where the contours include, respectively, 63.28\%, 95.45\% and 99.72\% of probability. From these posterior 2-D pdfs, we derive 1-D posterior pdfs for a few quantities. From their medians and $\pm 1$-$\sigma$ equivalent percentiles, we derive: $m_{\rm c} = 0.163^{+0.095}_{-0.020} \, \rm M_{\odot}$, $m_{\rm p} = 1.981^{+0.038}_{-0.088} \, \rm M_{\odot}$ and $M_{\rm T} = 2.157^{+0.028}_{-0.027} \, \rm M_{\odot}$. From the 2-$\sigma$ equivalent percentiles, we derive  $m_{\rm c} = 0.163^{+0.457}_{-0.023} \, \rm M_{\odot}$, $m_{\rm p} = 1.981^{+0.070}_{-0.409} \, \rm M_{\odot}$ and $M_{\rm T} = 2.157\pm0.055 \, \rm M_{\odot}$. Again, these values are consistent with those derived by \cite{Freire+2008}. The pulsar could still have a mass similar to the largest NS masses that have been reliably determined (PSR~J0348+0432, $m_{\rm p} = 2.01 \pm 0.04 \, \rm M_{\odot}$, \citealt{Antoniadis+2013} and PSR~J0740+6620, $m_{\rm p} = 1.99 \pm 0.07 \, \rm M_{\odot}$, \citealt{Fonseca+2021}), but not larger.

As we can see in the top left side panel of Fig.~\ref{fig:mass_mass}, there is no constraint on the orbital inclination, with the median of the $\cos i$ pdf appearing very close to 0.5. This indicates that, apart from the obvious fact that the Shapiro delay is not detectable (thus excluding orbital inclinations close to $90^\circ$), there are no additional detectable relativistic effects in the timing data. This means that, apart from the small likelihood of lower inclinations (which stems from their small range of $\cos i$), no measured relativistic effect precludes a substantially smaller pulsar mass and larger companion mass.

Given the fact that we are already timing M5 with FAST, and that 34 years have elapsed since discovery, the prospects for the determination of additional post-Keplerian parameters and individual mass measurements for this system appear poor in the foreseeable future. Given the low orbital eccentricities of the remaining binaries, the prospect of $\dot{\omega}$ measurements for them is even more distant.

\subsection{Detection of the Shapiro delay of M5F and possibly of M5D}\label{subs_3.5}
\label{sec:M5F}
If we use the orthometric parameterization of the Shapiro delay \citep{FreireWex2010} to fit for the Shapiro delay in M5F, we obtain a stable convergence, at $\varsigma = 0.982\, \pm \,0.043$ and $h_3 = 0.95 \pm 0.43 \, \rm \mu s$ for the orthometric ratio and amplitude, respectively. If we fix $\varsigma$ at this value, we obtain $h_3 = 0.95 \pm 0.34 \, \rm \mu s$, indicating a near 3-$\sigma$ detection of the Shapiro delay. This detection points to a very high inclination (close to $89^\circ$), $m_{\rm c} \, \sim \, 0.2 \, \rm M_{\odot}$ and $m_{\rm p} \, \sim \, 1.4 \, \rm M_{\odot}$, but still with large ($\sim 50\%$) relative uncertainties in $m_{\rm c}$. These values are consistent with the optical values derived below in section~\ref{sec:optical}, seen in particular in the rightmost plot of Fig.~\ref{fig:com-msps}. They are also consistent, within their wide uncertainties, with the $0.22 \rm \, M_{\odot}$ predicted by \cite{TS1999} for a Helium WD derived from Population II stars in a binary with the orbital period of M5F. A 1.5-$\sigma$ $h_3$ is also seen in the timing of M5D. An intense, well-coordinated FAST campaign might possibly measure the masses in the M5F system (and possibly in M5D) via Shapiro delay.

\subsection{Proper Motions}\label{subs_3.6}
Because of the increased timing baselines, the proper motions for M5A, M5B and M5C are significantly more precise than previous published values for these pulsars.
With timing baseline of $\sim$19 yr for M5D and E, we have also measured their proper motions. The smaller timing baselines for M5F and M5G, which are only detectable in FAST data, preclude a significant detection of the proper motion. The (unweighted) average proper motion for the first five pulsars in Right Ascension and Declination is $\mu_{\alpha}=4.14$\,mas yr$^{-1}$ and $\mu_{\delta}=-10.25$\,mas yr$^{-1}$, respectively. The standard deviations of the proper motions around this mean ($\sigma_{\mu}$) are 0.3\,mas yr$^{-1}$ in $\mu_{\alpha}$ and 0.89\,mas yr$^{-1}$ in $\mu_{\delta}$. The uncertainty in the mean value is given by $\sigma_{\mu}/\sqrt{N}$, where N is the number of measurements (5, in this case). Thus, our uncertainties for the mean cluster motion are $\sigma_{\mu, \alpha}=0.14$\,mas yr$^{-1}$ and $\sigma_{\mu, \delta}=0.4$\,mas yr$^{-1}$. At the distance to M5, these rms values translate to velocity dispersions of $\sim$5 and $\sim$14 km/s, which are of a similar order of magnitude to the stellar velocity dispersion ($\sigma_{\rm S} = 7.7 ~\rm km ~ s^{-1}$, \citealt{Baumgardt18, Lanzoni+2018, Kamann+2018}). However, the larger rms in declination is also likely to be caused, in part, by the larger uncertainties of the proper motions along that direction.

\begin{figure}[]
\centering 
\includegraphics[width=\columnwidth]{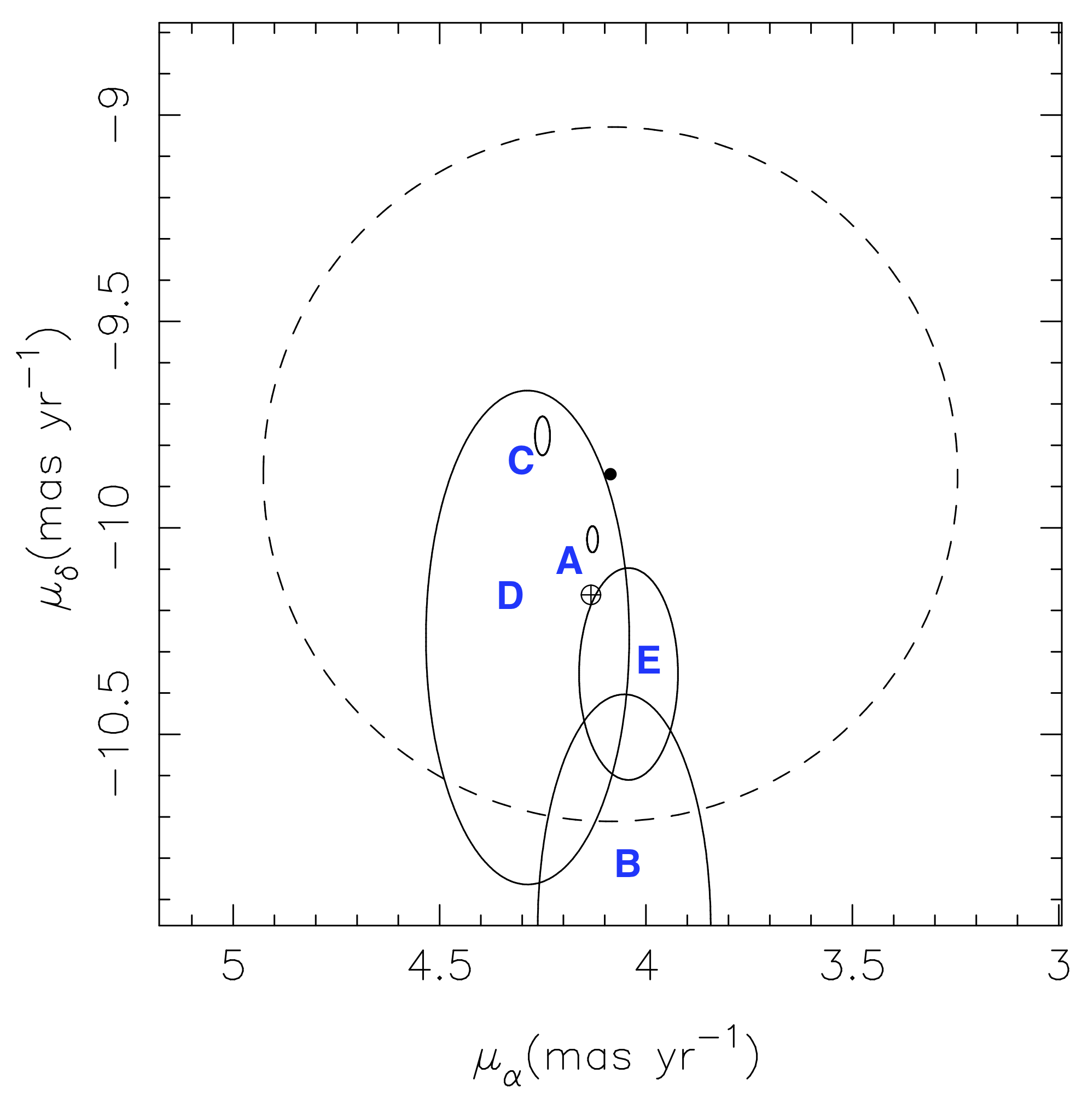}
\caption{Proper motions of the pulsars in M5. The error ellipses depict the 1-$\sigma$ uncertainties for their proper motion in $\alpha$ and $\delta$. The central dot is the proper motion derived from Gaia EDR3 measurements, the dashed circle around it depicts the escape velocity from the cluster. The Earth symbol ($\oplus$)} depicts the average of the pulsar proper motions.
\label{fig:pm}
\end{figure}

The proper motions of these five pulsars are depicted in Figure~\ref{fig:pm}, and the mean cluster motion along with the uncertainty is marked by the Earth symbol. According to the latest study based on the Gaia EDR3 data, \cite{Vasiliev21} derived M5's absolute proper motion as $\mu_{\alpha}=4.086\pm0.023$\,mas yr$^{-1}$ and $\mu_{\delta}=-9.870\pm0.023$\,mas yr$^{-1}$. This is consistent with our measurement of the average proper motion, if we take the latter's uncertainty into account. As we can see in Fig.~\ref{fig:pm}, all measured proper motions fall within the circle defined by the optical proper motion and the central escape velocity of M5, 29.9\,km/s~\citep{Baumgardt18}, with the exception of M5B. However, the latter still has a large uncertainty in its proper motion, its error ellipse overlaps with regions well within the escape velocity from the cluster. The best-measured proper motions (those of M5A and C) are the ones closer to the Gaia EDR3 proper motion.

\subsection{Acceleration in the cluster field}\label{subs_3.7}
In Table~\ref{tab:M5ephem}, we see that one of the pulsars (M5B) has a negative $\dot{P}$. Radio MSPs are rotationally powered, so their intrinsic spin period derivative $\dot{P}_{\rm int}$ should be positive.
The observed spin period derivative is given by \citep{Phinney+1993}:
\begin{equation}
\left( \frac{\dot{P}}{P} \right)_{\rm obs} = \left( \frac{\dot{P}}{P} \right)_{\rm int} + \frac{\mu^2d}{c} + \frac{a_{l,\, \rm GC}}{c} + \frac{a}{c},
\label{eq:pdot}
\end{equation}
where $\mu$ is the total proper motion of the system (for pulsars where this is not used, we used the proper motion of M5 discussed above), $d$ is the distance to the cluster ($\mu^2d / c$ is the so-called  Shklovskii effect; \citealt{Shklovskii1970}), $c$ is the speed of light, $a_{l,\, \rm GC}$ is the line-of-sight acceleration of the pulsar in the gravitational field of the cluster, and $a$ is the line-of-sight difference of the Galactic acceleration of the center of mass of M5 and that of the Solar System barycenter.
Using the \cite{McMillan2017} mass model of the Galaxy we obtain, for the position of M5, $a = -0.233 \times 10^{-9} \, \rm m \, s^{-2}$.

\begin{table*}[!htpb]
\centering
\caption{For each of the pulsars in M5, we calculate an upper limit for the pulsar accelerations in the field of the GC, theoretical upper and lower limits for the line-of-sight acceleration due to the cluster potential; the resulting limits on the intrinsic spin period derivative and extreme limits for the surface magnetic field strength ($B$) and the characteristic age ($\tau_c$) respectively (see text for details).
\label{tab:accelerations}}
\setlength{\tabcolsep}{4mm}{
\begin{tabular}{cccccc} 
\hline
Pulsar name & $a_{l, \rm P, max}$ & $a_{l, \rm GC, max}$ & $\dot{P}_{\rm int}$ & $B$ & $\tau_{\rm c}$ \\
            & $ 10^{-9} \rm m \, s^{-2}$ & $ 10^{-9}\, \rm  m \, s^{-2}$ &  $ 10^{-20} \, \rm s \, s^{-1}$ & $10^9$ G & Gyr  \\
\hline
A   & +1.82  &   1.83 &    0     - 6.76   &  0    - 0.62 &  1.3  - $\infty$   \\ 
B   & $-$0.64 &  1.84 &    0     - 3.20   &  0    - 0.51 &  3.9  - $\infty$ \\
C   & +2.76  &   3.03 &    0     - 4.80   &  0    - 0.35 &  0.8  - $\infty$ \\ 
D   & +1.78  &   1.60 &    0.19  - 3.37   &  0.08 - 0.32 &  1.4  - 25 \\
E   & +1.25  &   2.80 &    0     - 4.30   &  0    - 0.37 &  1.2  - $\infty$ \\
F   & +2.11  &   2.34 &    0     - 3.93   &  0    - 0.33 &  1.1  - $\infty$ \\
G   & +0.97  &   0.95 &    0.02  - 1.76   &  0.02 - 0.22 &  2.5  - 247 \\
\hline
\end{tabular}}
\end{table*}

\begin{figure}[]
\centering 
\includegraphics[width=\columnwidth]{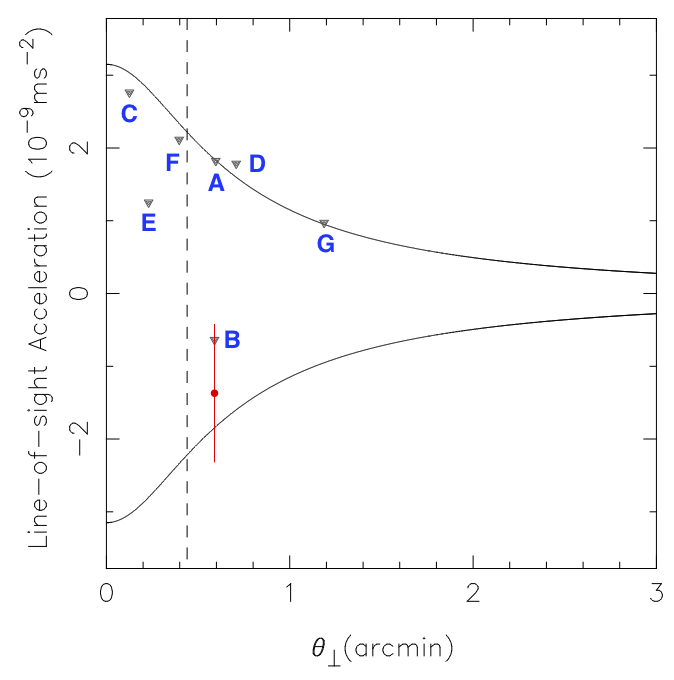}
\caption{Upper limits for the line-of-sight accelerations of the M5 pulsars (triangles) as a function of their angular distance from the nominal cluster center ($\theta_{\perp}$), derived from $\dot{P}_{\rm obs}$. The solid black curves indicate the maximum and minimum accelerations along the line of sight caused by the potential of the globular cluster predicted by the analytical model described in the text. The red dot and its error bar indicate the acceleration (and its uncertainty) derived from the orbital period derivative of M5B.
The vertical dashed line indicates the core radius.}
\label{fig:pulsar_accelerations}
\end{figure}

The acceleration caused by the gravitational potential of the GC, $a_{l, \rm GC}$, is generally the dominant contribution. To model this, we used an analytical model of the cluster described in \cite{Freire+2005}, which is based on the empirical \cite{King1962} density profile. The line-of-sight acceleration due to the cluster potential at distance $x$ from the centre (in core radii) and distance $l$ from the plane of the sky passing through the centre of the GC (also in core radii) is given by:
\begin{equation}
a_{l, \rm GC}(x) = \frac{9 \sigma_{\rm S}^2}{d \theta_c} \frac{l}{x^3}
\left( \frac{x}{\sqrt{1+x^2}}  - \sinh^{-1}x \right).
\end{equation}
The parameters for this model include the position, distance and core radius of M5 (see Introduction) and the aforementioned central stellar velocity dispersion. In Fig.~\ref{fig:pulsar_accelerations}, the solid black curves represent the maximum and minimum values of $a_{l, \rm GC}(x)$ for each angular offset from the centre, $\theta_{\perp}$.

For each pulsar, if we assume a negligible intrinsic spin-down, we can derive an absolute upper limit for its acceleration in the field of the GC:
\begin{equation}
a_{l, \rm P, max} \, = \, c \frac{\dot{P}_{\rm obs}}{P} - \mu^2 d - a,
\end{equation}
where, if the pulsar does not have a well-measured proper motion, we used the proper motion for the cluster measured by \cite{Vasiliev21}. The upper limits appear in Fig.~\ref{fig:pulsar_accelerations} as the triangles; the values of these accelerations are also listed in Table~\ref{tab:accelerations}.

From this figure, we conclude that, despite the small accelerations predicted by the analytical model described above, it can account for the negative $\dot{P}_{\rm obs}$ of M5B (for the line-of-sight acceleration to be negative, this binary must be located in the more distant half of the cluster). The model cannot fully account for the positive $ a_{l, \rm P, max}$ of pulsars M5D and G, but this is to be expected because their $\dot{P}_{\rm obs}$ has a contribution from a positive $\dot{P}_{\rm int}$. 

Taking the maximum and minimum theoretical accelerations caused by the gravitational field of the GC for the line of sight of each pulsar, we can calculate maximum (and in some cases minimum) limits for the $\dot{P}_{\rm int}$ of each pulsar; from these we can derive extreme limits for their magnetic fields and characteristic ages. These values are also presented in Table~\ref{tab:accelerations}.

An estimate of the intrinsic spin-down can be also derived from the observed orbital period derivative of M5B. Given its $P_{\rm b}$ of 6.85 days, low eccentricity and small companion mass, the intrinsic variation caused by GW emission is negligible, so any variations of $P_{\rm b}$ are caused by the three last terms in an equation similar to Eq.~\ref{eq:pdot}. Subtracting the latter equation from Eq.~\ref{eq:pdot}, we obtain
\begin{equation}
 \left( \frac{\dot{P}}{P} \right)_{\rm int} =
\left( \frac{\dot{P}}{P} \right)_{\rm obs} -  \left( \frac{\dot{P}_{\rm b}}{P_{\rm b}} \right)_{\rm obs},
\label{eq:pdot2}
\end{equation}
from which we obtain $\dot{P}_{\rm int} = (1.9 \pm 2.5) \times 10^{-20}$. From the nominal value we obtain  $B_0 \sim 4 \times 10^{8} \, \rm G$ and
$\tau_c \sim 6.5 \, \rm Gyr$; however, given the large relative uncertainty in $\dot{P}_{\rm int}$, these values are still very crude approximations.

The fact that the majority of pulsars has positive $\dot{P}_{\rm obs}$ suggests that the positive $\dot{P}_{\rm int}$ are of a similar magnitude to the effect of the GC acceleration, otherwise about half of pulsars would have negative $\dot{P}_{\rm obs}$, as observed for GCs with much larger predicted accelerations, like 47~Tuc \citep{Freire+2017,Abbate+18} or Terzan 5 \citep{Prager+2017}. However, this is still small number statistics: even if all $\dot{P}_{\rm int}$ were very small compared to the accelerations, there would still be a 5\% probability (using \citealt{Gehrels86}, Table 6) that only one out of 7 has a negative acceleration.

The main conclusion to be taken from this analysis is that, even though the individual accelerations in the gravitational field of M5 are not known for most pulsars, the predicted accelerations have a rather small range. From this and the measured $\dot{P}_{\rm obs}$, we conclude that $B_0 < 6.2 \times 10^{8}$G and $\tau_c > 0.8$Gyr. This is consistent with the characteristics of the MSP population in the Galactic disk.

\subsection{Optical counterparts to MSPs}\label{subs_3.8}

\label{sec:optical}

\begin{figure*}[]
    \centering
    \includegraphics[width=4.5cm]{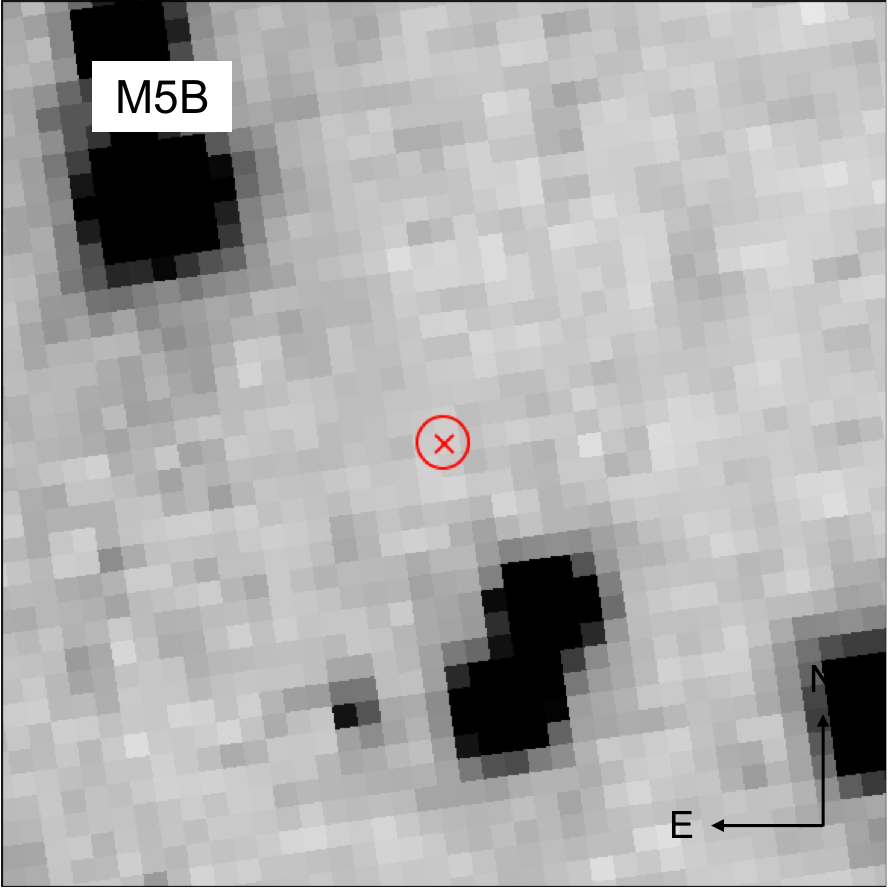}
    \includegraphics[width=4.5cm]{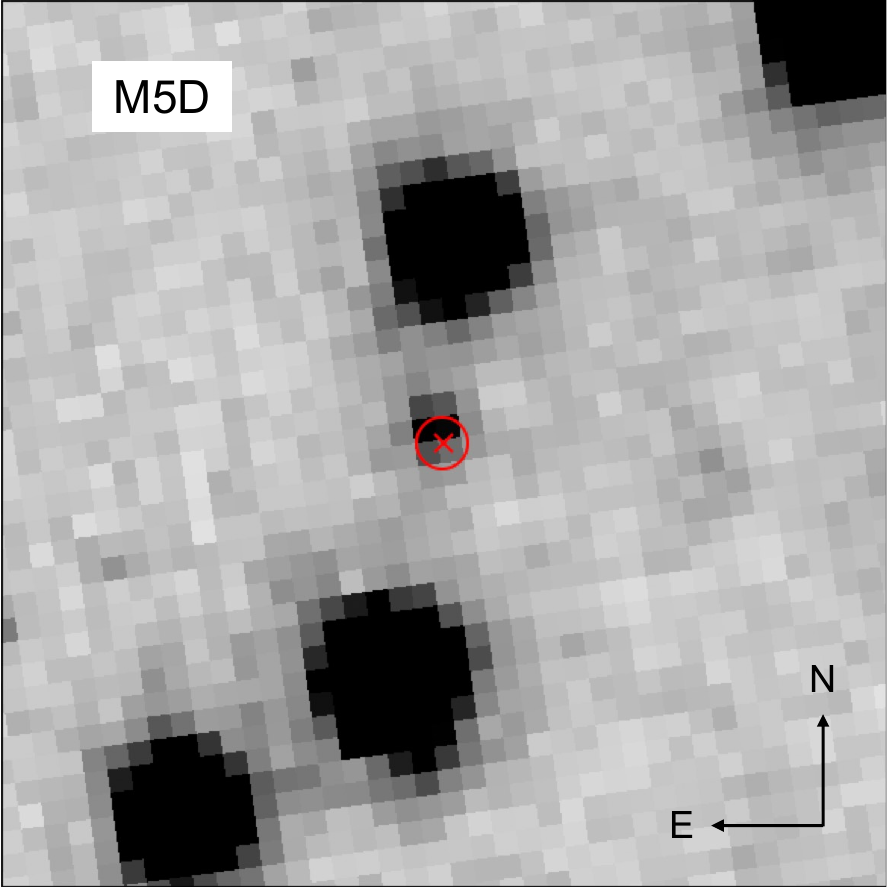}
    \includegraphics[width=4.5cm]{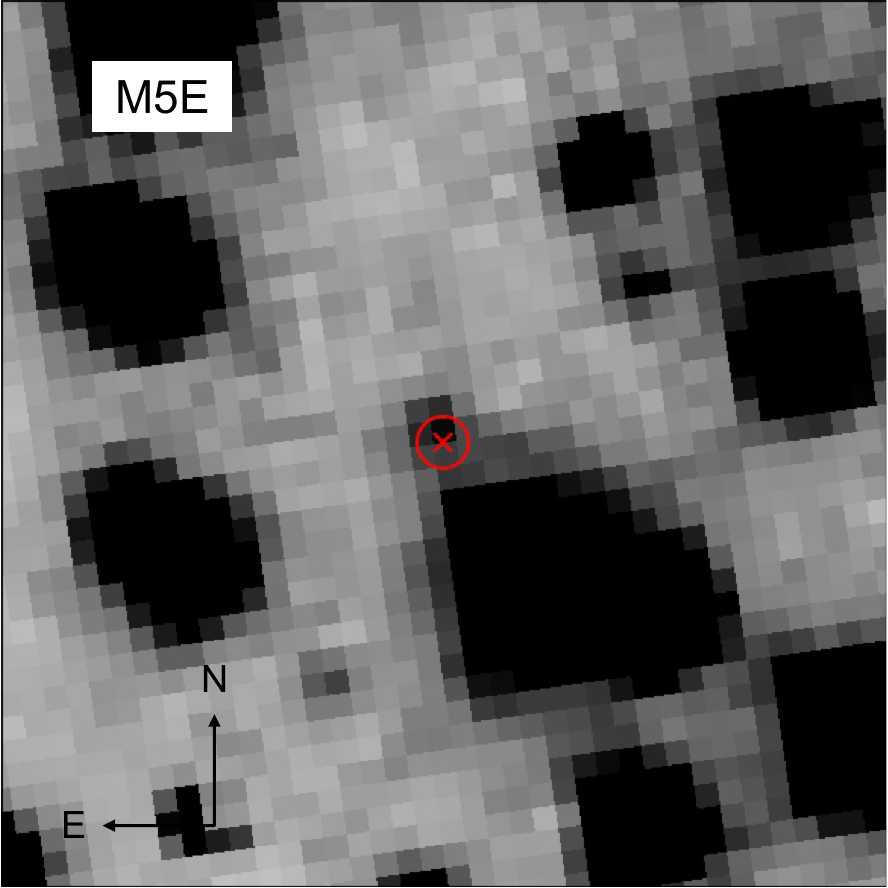}
    \includegraphics[width=4.5cm]{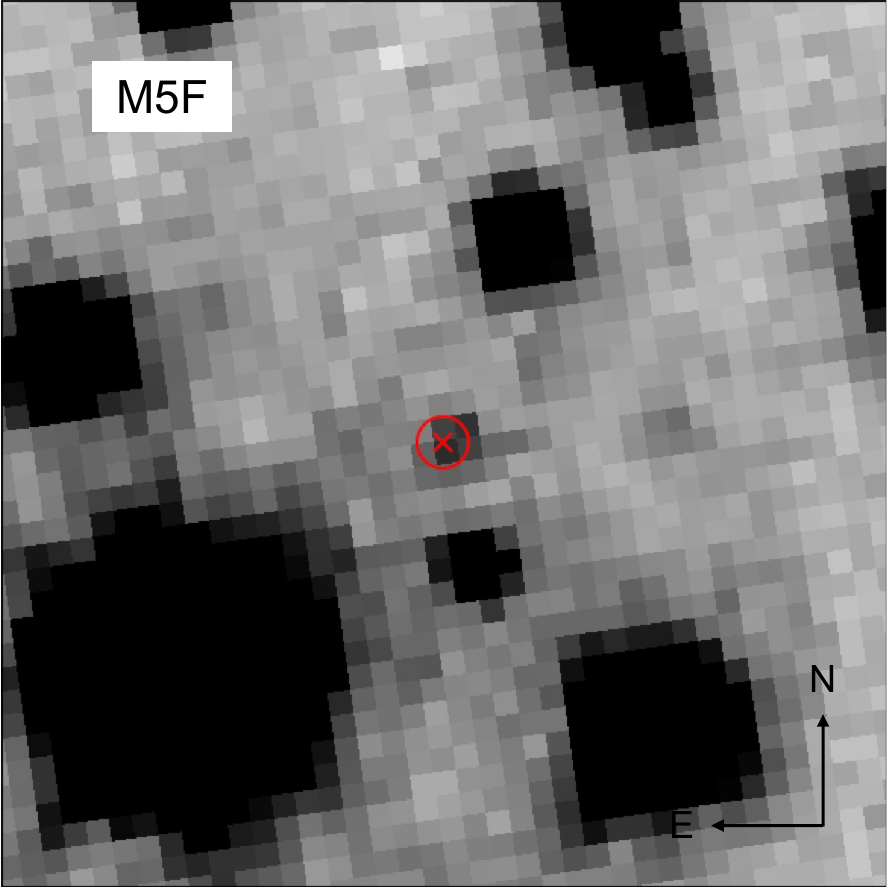}
    \includegraphics[width=4.5cm]{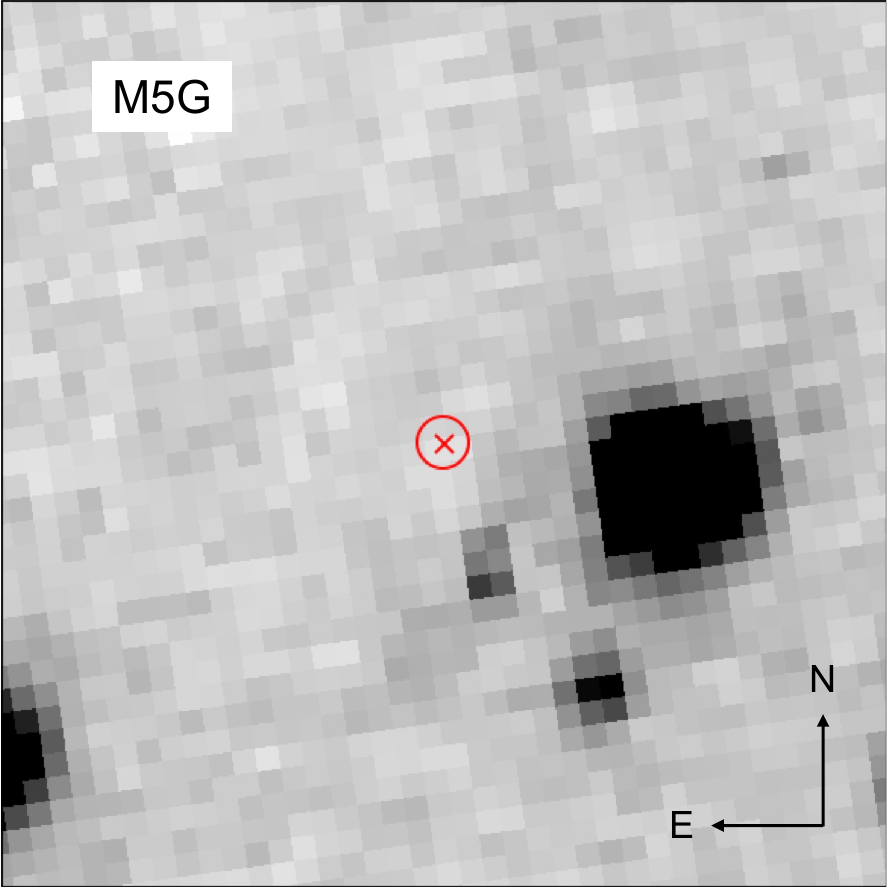}
    \caption{$1.5\arcsec \times 1.5\arcsec$ finding charts of the regions surrounding the positions of M5B, M5D, M5E, M5F, and M5G in a combined F275W image. In each panel, the red cross indicates the pulsar position and the red circle has a radius of $3\sigma$, using the combined optical and radio positional uncertainties.
    }
    
    \label{fig:fc}
\end{figure*}
In previous works, two of the binary MSPs in M5 have been studied at optical wavelengths. The companion of M5B is not detectable \citep{Freire+2008}, a result that was important to establish that the observed periastron advance is relativistic, which we confirm below. However, the low-mass companion of the M5C ``black widow" system 
has been detected 
\citep{Pallanca+2014}, showing the photometric variability with orbital phase that is characteristic of such systems.

\begin{figure*}
    \centering
    \includegraphics[width=0.8\textwidth]{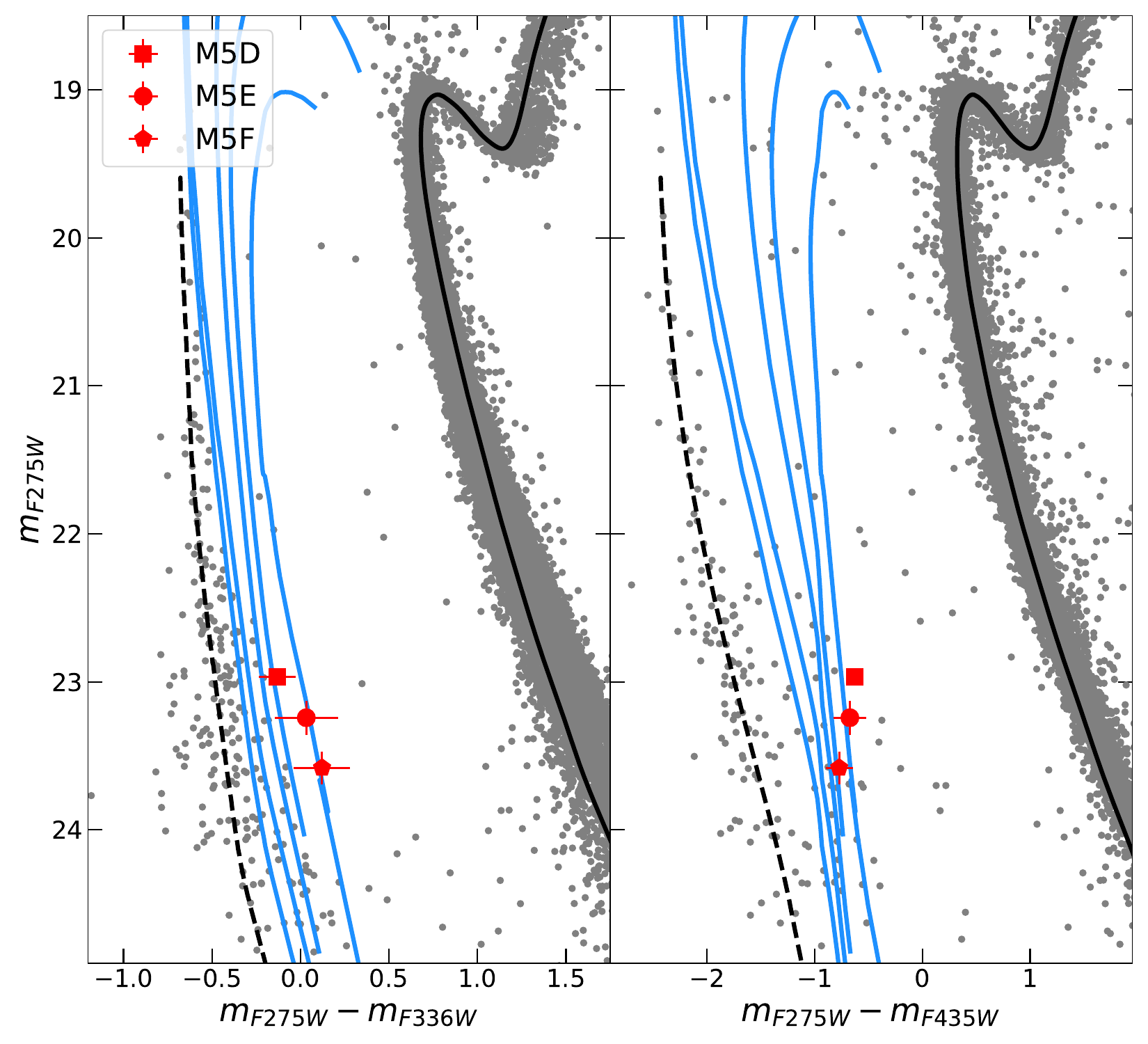}
    \caption{{\it Left-hand panel:} CMD of M5 in a combination of the F275W and F336W filters. The red square, circle and pentagon are the positions of the counterparts to M5D, M5E and M5F, respectively. The solid and dashed black curves are a 12.5 Gyr stellar population isochrone and a $0.55 M_{\odot}$ white dwarf cooling track, respectively. Blue curves are cooling tracks of He-WDs with masses of  0.17 $M_{\odot}$, 0.19 $M_{\odot}$, 0.21 $M_{\odot}$, 0.25 $M_{\odot}$ and 0.30 $M_{\odot}$, with increasing masses from left to right. {\it Right-hand panel}: Same as in the left-hand panel, but with a combination of F275W and F435W filters.}
    \label{fig:cmd}
\end{figure*}

In order to identify the undetected optical counterparts to the remaining binary MSPs, we investigated all stars located around their positions. At the corresponding positions of M5D, M5E and M5F we found three blue objects that in the color-magnitude diagrams (CMDs) are located along the red side of the white dwarf cooling sequence in all the available filter combinations (see an example in Figure~\ref{fig:cmd}). This is the CMD position where He-core white dwarf (He-WD), i.e. the typical outcome of the evolution of binary MSPs, are located. All the companions are located at distances between 50 mas and 60 mas from the corresponding pulsar positions. Such distances are larger than the combined optical and radio position uncertainties. However, the three candidate counterparts are systematically shifted toward the northwest direction, suggesting an offset between the radio and optical frames. The average shift along RA and Dec is $-0.03\arcsec$ and $0.05\arcsec$, respectively. After accounting for such a shift, all the three white dwarfs are located at distances between 4 mas and 15 mas from the pulsar positions. We detected no stars at the corresponding positions of M5B and M5G. The finding charts of the 5 investigated objects, after accounting for the radio-optical offset, are presented in Figure~\ref{fig:fc}. Magnitudes and upper limits are listed in Table~\ref{tab:magnitudes}.

To confirm the He-WD nature of the three counterparts and derive their physical properties we compared their magnitudes obtained in different filters with those predicted by binary evolution models. First, we performed a photometric calibration sanity check in all the available filter combinations by comparing the standard evolutionary sequences with a 12.5 Gyr isochrone extracted from the BaSTI database 
\citep{2018ApJ...856..125H, 2021ApJ...908..102P}
assuming $[Fe/H]=-1.3$ \citep[][2010 edition]{1996AJ....112.1487H}.  We also compared the position of the observed white dwarf cooling sequence with a theoretical cooling track extracted from the BaSTI database 
\citep{2022MNRAS.509.5197S}
for CO white dwarfs with a typical mass of $0.55 M_{\odot}$. Absolute magnitudes were converted to the observed frame assuming a distance modulus $(m-M)_0=14.37$ and a color excess $E(B-V)=0.03$ \citep[][2010 edition]{1996AJ....112.1487H} and using appropriate extinction coefficients from  \citet{1989ApJ...345..245C} and \citet{1994ApJ...422..158O}.
All the evolutionary sequences are well reproduced by the adopted models in all the filter combinations (see 
the black curves in Figure~\ref{fig:cmd}.)

In order to constrain the properties of the companion stars, we exploit the binary evolution models from the database described in 
\citet{2014A&A...571L...3I, 2016A&A...595A..35I}. Theoretical luminosities and temperatures were converted to observed magnitudes following the prescriptions by \citet[see also \citealt{2019ApJ...875...25C}]{2023ApJ...948...84C}. A selection of the evolutionary tracks is plotted in Figure~\ref{fig:cmd}.
It can be clearly seen that the positions of the counterparts are compatible with those expected by the evolution of low-mass He-core white dwarfs. To get a quantitative estimate of the companion physical properties (such as their masses, radii, cooling ages, surface gravities, and temperatures) we implemented the same approach described in detail in 
\citet[see also \citealt{2020ApJ...905...63C}]{2019ApJ...875...25C}, by defining a logarithmic likelihood (see Equation~1 in \citealt{2020ApJ...905...63C}) to quantify the probability of each point of each evolutionary track to reproduce the observed companion magnitudes in all the available filters. Then, for each of the investigated parameters, we obtained an estimate and its related uncertainty as the 0.16, 0.5, and 0.84 quantiles of the likelihood distributions. The derived properties of the three companions are listed in Table~\ref{tab:opt}. Notably, all the three companions are extremely low-mass white dwarfs with similar masses, as commonly found for similar objects in other GCs. 

\begin{figure*}
    \centering
    \includegraphics[width=\textwidth]{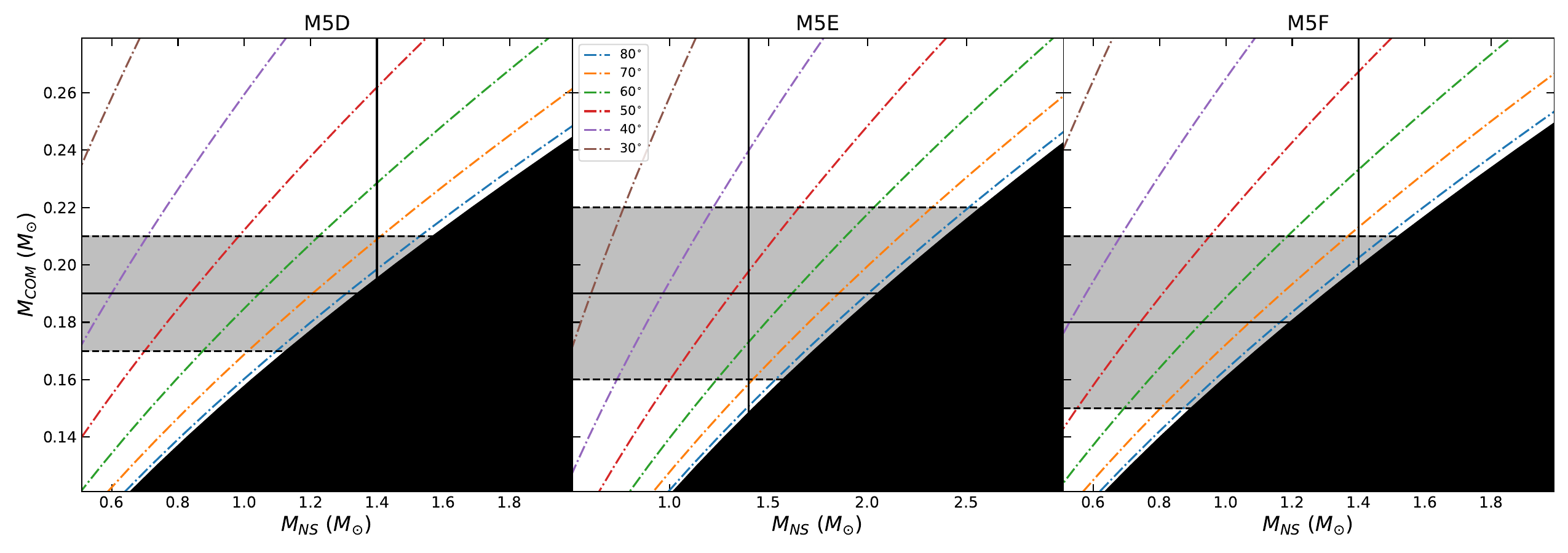}
    \caption{The mass of the neutron star as a function of companion mass  for M5D, M5E, and M5F. The derived mass of the companion star is marked with a horizontal solid line, and the corresponding uncertainty is limited by two horizontal dashed lines in shallow areas. The colored curves represent the relationship between the companion masses and the neutron star masses at different inclinations and the black region is excluded by the pulsar mass function.}
    \label{fig:com-msps}
\end{figure*}

The determination of the companion masses coupled with the binary orbital parameters (see Table~\ref{tab:opt}) can be used to constrain the orbital inclination angles and possibly, the neutron star masses. In fact, the masses of the binary components can be expressed as a function of the orbital parameters through the mass-function. Figure~\ref{fig:com-msps} shows, for each of the three investigated systems, the companion mass as a function of the neutron star mass. Different curves correspond to different inclination angles, as predicted by the mass-functions. In the case of M5D and M5F, it is most likely that the two binaries host a canonical neutron star. In fact, assuming a standard neutron star mass of $1.4 M_{\odot}$, it can be seen that the derived companion masses imply that both the systems are likely observed at very high inclination angles ($i\geq80^{\circ}$);  which is consistent with the detection of the Shapiro delay of M5F and, marginally, of M5D (section~\ref{sec:M5F}). On the other hand, in the case of M5E, it is not possible to rule out the presence of a more massive neutron star. 
M5E could host a canonical neutron star and be observed at intermediate-high inclination angles ($45^{\circ}<i<70^{\circ}$) or, alternatively, could host a more massive neutron star at high inclination angles.

\begin{table*}[htbp]
    \centering
    \begin{threeparttable}
    \caption{Magnitudes and magnitude upper limits measured in each available filter for the 5 investigated MSPs.}
    \begin{tabular}{c c c c c c c} \hline \hline
       Name  & F275W            & F336W           & F390W           & F435W           & F606W            &  F814W           \\ \hline
    COM-M5B  & $>$ 24.87        & $>$ 25.24       & $>$ 25.85       & $>$ 25.98       & $>$ 25.43        & $>$ 23.48        \\
    COM-M5D  & 22.97$\pm$0.05   & 23.10$\pm$0.09  & 23.70$\pm$0.02  & 23.59$\pm$0.03  & 23.57$\pm$0.02   & 23.08$\pm$0.20   \\
    COM-M5E  & 23.24$\pm$0.12   & 23.21$\pm$0.14  & 23.95$\pm$0.05  & 23.92$\pm$0.10  & -                & -                \\
    COM-M5F  & 23.58$\pm$0.11   & 23.46$\pm$0.12  & 23.82$\pm$0.05  & 24.35$\pm$0.06  & 23.85$\pm$0.08   & 23.4$\pm$0.20    \\
    COM-M5G  & $>$ 24.41        & $>$ 24.11       & $>$ 25.13       & $>$ 25.18       & $>$ 25.35        & $>$ 25.01        \\
    \hline 
    \end{tabular}
    \label{tab:magnitudes}
    \begin{tablenotes}
        \item {\bf Note:} The companion star to M5E is heavily contaminated by nearby bright stars in the F606W and F814W, thus no magnitudes or upper limits can be derived.
    \end{tablenotes}
    \end{threeparttable}
\end{table*}

\begin{table}[]
\centering
\caption{Derived properties of the companions to M5D, M5E, and M5F.}
\scriptsize
\setlength{\tabcolsep}{0.25mm}{
\begin{tabular}{l c c c} \hline
                                              &    M5D  &  M5E  &  M5F  \\ \hline
Companion mass, $m_{\rm c}$  ($\rm M_\odot$)  & $0.19\pm0.02$     &   $0.19\pm0.03$      &  $0.18\pm0.03$  \\
Surface gravity, log $g$ (cm s$^{-2}$) &  $6.4\pm0.3$   &   $6.5\pm0.3$    &   $6.3\pm0.3$    \\
Companion radius, $R_{\rm c}$  ($\rm R_\odot$)  & $0.044\pm0.01$     &  $0.042\pm0.01$    &  $0.051\pm0.015$ \\
Effective temperature, $T_{\rm eff}$  (K)    &  $10400^{+1400}_{-800}$  & $10000^{+1300}_{-1000}$  &  $8800^{+1100}_{-1000}$    \\
Cooling age (Gyr)       &   $2.3\pm1.0$     &   $2.8\pm1.4$    &  $1.8^{+2.0}_{-0.5}$     \\ \hline
\multicolumn{4}{p{0.9\columnwidth}}{{\bf Note}: Companion masses are from optical analysis.} \\
\end{tabular}}
\label{tab:opt}
\end{table}

\subsection{X-ray counterparts to MSPs}\label{subs_3.9}

We performed X-ray data reduction and analysis using {\sc ciao}\footnote{Chandra Interactive Analysis of Observations, available at \url{https://cxc.harvard.edu/ciao/}.} \citep[version 4.15.1 with {\sc caldb 4.10.2}]{Fruscione2006}. 
We first reprocessed the dataset to create a new level=2 event file and a new bad pixel file using {\tt chandra\_repro} script. 
We performed source detection in the 0.3--8 keV image using the {\tt wavdetect} script\footnote{\url{https://cxc.cfa.harvard.edu/ciao/threads/wavdetect/}}, with a scale list of [1, 1.4, 2, 4, 8], and a significance threshold of 10$^{-6}$. 
Four X-ray sources were detected in the vicinities of MSPs C, D, E, and G, respectively (Table~\ref{tab:x-ray_cp}), while no X-ray sources were detected around the timing positions of MSPs A, B, and F (see the zoom-in images in Figure~\ref{fig:chandra_m5}). 
Therefore, for the counterparts to MSPs C, D, E, and G, we were able to extract their X-ray spectra and conduct rigorous spectral analysis. 
We note that we reanalysed the X-ray spectrum of M5C to keep consistency in this work, though it has been presented in \citet{Zhao+2022}. 

We applied {\tt specextract} script to extract their spectra in 1-arcsec-radius regions centered at the corresponding X-ray positions (see Table~\ref{tab:x-ray_cp}), while the background spectra were extracted from nearby source-free regions. 
We then used the spectral analysis software {\sc bxa} \citep{Buchner2014}, which connects the nested sampling algorithm UltraNest \citep{Buchner2021} with {\sc sherpa}\footnote{{\sc ciao}'s modelling and fitting package, available at \url{https://cxc.cfa.harvard.edu/sherpa/}.}, to conduct Bayesian parameter estimation and model comparison. 
We fitted their spectra with three absorbed spectral models; blackbody (BB), neutron star hydrogen atmosphere \citep[NSA;][]{Heinke2006}, and power-law (PL), realized by {\tt xsbbodyrad}, {\tt xsnsatmos}, and {\tt xspegpwrlw}, respectively, in {\sc sherpa}, which are commonly observed spectra from MSPs \citep[see e.g.][]{Zhao+2022}. 
The interstellar absorption towards M5 was modelled by the {\tt xstbabs} model, with {\it wilm} abundances \citep{Wilms2000} and {\it vern} photoionization cross sections \citep{Verner1996}, while the hydrogen column density ($N_{\rm H}$) was assumed to be fixed at 2.61$\times$10$^{20}$ cm$^{-2}$. 
This $N_{\rm H}$ was estimated using the correlation between $N_{\rm H}$ and optical extinction ($A_V$) in \citet{Bahramian2015}, while the $A_V$ towards M5 was calculated as $A_V=3.1 \times E(B-V)$ \citep{Cardelli1989}, where $E(B-V)=0.03$ is the foreground reddening towards M5 \citep{Harris+2010}. 

To generate Bayesian parameter estimation, we defined uniform priors for the PL photon index ($\Gamma$) between 0 and 5, for NSA $\log_{10}T_{\rm eff}$ between 5 and 6.5, and a log-uniform prior for BB $kT_{\rm BB}$ between 10$^{-4}$ and 10, where $T_{\rm eff}$ is the unredshifted effective temperature in units of Kelvin and $kT_{\rm BB}$ is the BB temperature in keV. In addition, for the NSA model, the neutron star mass and radius were set to be 1.4~M$_\sun$ and 10 km, respectively, and the distance to M5 was assumed to be 7.5 kpc \citep{Harris+2010}. All priors for normalization parameters were defined to be log-uniform between 10$^{-8}$ and 10$^2$. Note that the normalization parameters of BB and NSA models can be used to infer the effective surface emitting regions ($R_{\rm eff}$), whereas the normalization of the PL model directly returns the X-ray flux in a given band. Each X-ray spectrum was grouped to at least one photon per bin and filtered to 0.3--8 keV range in the fitting process. We used the difference of log Bayesian evidences ($Z$) from the fitting results to compare models, where we adopted a difference of $\Delta\log_{10}Z>1.5$ to rule out models \citep[i.e. Jeffreys' scale; see][]{Jeffreys1939}.

We present the spectral fitting results and X-ray properties of the four X-ray counterparts in Table~\ref{tab:x-ray_fits}. We found that all spectral models returned acceptable fits, with the NSA model having the largest likelihoods, whereas we were not able to rule out any spectral models by simply comparing the Bayesian evidences. 
On the other hand, the photon indices obtained from the PL model were all greater than $\sim$2.5, implying substantial soft, thermal X-ray emission from these sources, which favors blackbody-like spectral models, BB and NSA. Moreover, the best-fit BB and NSA parameters are similar to those typically observed from MSPs in GCs \citep[e.g.][]{Bogdanov2006}. It is also noticeable that the NSA model returned a likelihood about 10 times larger than the BB model in each of the four spectral fittings, suggesting that the NSA model better represents the thermal emission from MSPs, though both models are acceptable. Additionally, we examined combined models, such as PL+BB and PL+NSA, in spectral fittings, but these combined models did not improve fits significantly, and the parameters were more poorly constrained. Hence, based on the X-ray spectral fits and properties, we identified X-ray counterparts to four MSPs in M5 (including three newly identified counterparts to MSPs D, E, and G, respectively), whose spectra are well fitted by an absorbed NSA model with X-ray luminosities between 6.0$\times$10$^{30}$~erg~s$^{-1}$ and 1.2$\times$10$^{31}$~erg~s$^{-1}$. We note that these X-ray luminosities are consistent with those typically observed for MSPs in GCs \citep[see e.g.][]{Zhao+2022}. However, sources in M5 with X-ray luminosities lower than $\sim$5$\times$10$^{30}$~erg~s$^{-1}$ \citep{Bahramian2020} are undetectable due to the long distance to this cluster and the limited exposure time (e.g. likely cases for MSPs A, B, and F). 

Due to the lack of source detections around MSPs A, B, and F, we were unable to perform stringent spectral analysis for them. Nonetheless, to constrain their X-ray luminosities, we extracted the spectra from 1.5-arcsec-radius regions centered at the timing positions of MSPs A, B, and F, respectively, which took into account possible radio and X-ray positional offsets. We then simply fitted their spectra with an absorbed PL model with a photon index fixed to 2. The choice of $\Gamma=2$ can include both thermal and non-thermal emission from an MSP and therefore reduce bias of model selection \citep{Bogdanov2021}. We also fitted their spectra with an absorbed NSA model by fixing $\log_{10}T_{\rm eff}=6$, which is the average of the four best-fit NSA parameters of MSPs C, D, E, and G. The fitting results of MSPs A, B, and F are also presented in Table~\ref{tab:x-ray_fits}. We found that the X-ray luminosities fitted by PL models are lower than $\sim$7$\times$10$^{30}$~erg~s$^{-1}$, while the luminosities obtained from NSA models are lower than $\sim$3.5$\times$10$^{30}$~erg~s$^{-1}$.
The difference in X-ray luminosities between the PL and NSA models is due to their different modeling of fluxes in the hard X-ray band, where nominal thermal X-rays are not expected. Considering the limiting X-ray luminosity of a source to be detected in M5 is about 5$\times$10$^{30}$~erg~s$^{-1}$ \citep{Bahramian2020}, the upper limits set by PL models are very conservative. 

\begin{table}
    \centering
    \caption{X-ray counterparts to four MSPs}
    \begin{tabular}{lccc}
\hline
MSP	&	X-ray position$^a$		&	PU$^b$	&	Offset$^c$	\\
Name & (hh:mm:ss.ss+dd:mm:ss.s) & (\arcsec) & (\arcsec) \\
\hline
C	&	15:18:32.81	$+$02:04:47.5	&	0.52	&	0.40	\\
D	&	15:18:30.40	$+$02:05:00.2	&	0.68	&	0.76	\\
E	&	15:18:33.27	$+$02:04:37.4	&	0.37	&	0.78	\\
G	&	15:18:28.73	$+$02:05:15.1	&	0.86	&	0.35	\\
\hline
\multicolumn{4}{p{0.9\columnwidth}}{{\bf Note}: X-ray detections of four MSP counterparts using {\tt wavdetect}.} \\
\multicolumn{4}{p{0.9\columnwidth}}{$^a$ The X-ray centroids reported by {\tt wavdetect}.} \\
\multicolumn{4}{p{0.9\columnwidth}}{$^b$ X-ray positional uncertainties at a 95\% confidence level, computed using Equation~12 in \citet{Kim+2007}.} \\
\multicolumn{4}{p{0.9\columnwidth}}{$^c$ Angular offsets between timing and X-ray positions.}
    \end{tabular}
    \label{tab:x-ray_cp}
\end{table}

\begin{table*}
    \centering
    \caption{X-ray spectral fitting results of M5 MSPs.}
    \begin{tabular}{lccccccc}
\hline
MSP	&	Spectral$^a$	&	Parameter$^b$ & {$R_{\rm eff}$}$^c$	&	{$F_{0.5-2}$}$^d$	&	{$F_{2-8}$}$^d$	&	{$L_{0.5-10}$}$^e$	&	{log$_{10}Z$}$^f$	\\
Name & Model & & (km) & \multicolumn{2}{c}{(10$^{-16}$~erg~cm$^{-2}$~s$^{-1}$)} & (10$^{30}$~erg~s$^{-1}$) & \\
\hline
A	&	PL	&	2	& -- &	$<$4.6	&	$<$4.6	&	$<$6.6	&	--	\\
 & NSA & 6 & $<$1.1 & $<3.8$ & $<$0.1 & $<$2.5 & --  \\
B	&	PL	&	2	& -- &	$<$4.7	&	$<$5.5	&	$<$6.8	&	--	\\
 & NSA & 6 & $<$1.3 & $<$4.8 & $<$0.1 & $<$3.2 & -- \\
C	&	BB	&	0.13$^{+0.03}_{-0.02}$	& 0.62$^{+0.41}_{-0.27}$ &	10.0$^{+4.0}_{-2.9}$	&	0.0$^{+0.0}_{-0.0}$	&	6.7$^{+2.7}_{-2.0}$	&	$-$1.1	\\
	&	NSA	&	5.80$^{+0.16}_{-0.16}$	& 6.00$^{+11.25}_{-3.66}$ &	9.7$^{+3.7}_{-3.0}$	&	0.0$^{+0.1}_{-0.0}$	&	6.5$^{+2.6}_{-2.0}$	&	0	\\
	&	PL	&	3.84$^{+0.74}_{-0.70}$	& -- &	8.6$^{+3.4}_{-2.5}$	&	0.7$^{+1.3}_{-0.5}$	&	6.3$^{+3.0}_{-2.0}$	&	$-$0.3	\\
D	&	BB	&	0.14$^{+0.04}_{-0.03}$	& 0.54$^{+0.49}_{-0.25}$ &	9.4$^{+4.4}_{-2.9}$	&	0.0$^{+0.0}_{-0.0}$	&	6.3$^{+3.0}_{-2.0}$	&	$-$0.7	\\
	&	NSA	&	5.83$^{+0.18}_{-0.17}$	& 4.83$^{+9.66}_{-3.18}$ &	8.8$^{+4.0}_{-2.9}$	&	0.0$^{+0.1}_{-0.0}$	&	6.0$^{+2.8}_{-2.0}$	&	0	\\
	&	PL	&	3.65$^{+0.79}_{-0.80}$	& -- &	8.3$^{+3.5}_{-2.8}$	&	0.9$^{+2.1}_{-0.6}$	&	6.4$^{+3.3}_{-2.4}$	&	$-$0.4	\\
E	&	BB	&	0.27$^{+0.06}_{-0.05}$	& 0.14$^{+0.07}_{-0.05}$ &	16.1$^{+3.5}_{-3.3}$	&	1.3$^{+1.7}_{-0.9}$	&	11.9$^{+2.9}_{-2.5}$	&	$-$0.9	\\
	&	NSA	&	6.29$^{+0.11}_{-0.12}$	& 0.50$^{+0.40}_{-0.20}$ &	15.8$^{+4.1}_{-3.4}$	&	1.7$^{+1.6}_{-0.9}$	&	11.9$^{+3.2}_{-2.8}$	&	0	\\
	&	PL	&	2.39$^{+0.44}_{-0.42}$	& -- &	13.9$^{+3.6}_{-2.9}$	&	8.2$^{+6.7}_{-4.1}$	&	15.9$^{+6.5}_{-4.3}$	&	$-$0.5	\\
F	&	PL	&	2	& -- &	$<$2.4	&	$<$2.4	&	$<$3.4	&	--	\\
 & NSA & 6 & $<$0.9 & $<$2.3 & $<$0.1 & $<$1.6 & -- \\
G	&	BB	&	0.23$^{+0.08}_{-0.05}$	& 0.15$^{+0.13}_{-0.07}$ &	8.7$^{+3.9}_{-2.8}$	&	0.3$^{+0.9}_{-0.2}$	&	6.3$^{+2.9}_{-2.2}$	&	$-$1.0	\\
	&	NSA	&	6.15$^{+0.18}_{-0.18}$	& 0.75$^{+1.32}_{-0.44}$ &	8.8$^{+3.5}_{-3.0}$	&	0.4$^{+0.9}_{-0.3}$	&	6.2$^{+2.7}_{-2.1}$	&	0	\\
	&	PL	&	2.75$^{+0.77}_{-0.65}$	& -- &	8.1$^{+2.9}_{-2.6}$	&	2.7$^{+4.7}_{-1.9}$	&	7.7$^{+5.0}_{-2.9}$	&	$-$0.2	\\
 \hline
 \multicolumn{8}{p{13.5cm}}{{\it Notes}: Due to the lack of detections of X-ray counterparts to MSPs A, B, and F, their spectra were only fitted with an absorbed power-law model by fixing $\Gamma=2$, and an absorbed NSA model by fixing $\log_{10}T_{\rm eff}=6$. The fitting results are presented as 1-$\sigma$ upper limits. $N_{\rm H}$ towards M5 was fixed at 2.61$\times$10$^{20}$~cm$^{-2}$ for all the fits. The quoted uncertainties represent 1-$\sigma$ confidence level. } \\
 \multicolumn{8}{p{13.5cm}}{$^a$ X-ray spectral models applied in this work. BB: blackbody; NSA; neutron star hydrogen atmosphere \citep{Heinke2006}; PL: power-law.} \\
 \multicolumn{8}{p{13.5cm}}{$^b$ The best-fit $kT_{\rm BB}$ (in keV), $\log_{10}T_{\rm eff}$ (in $\log_{10}{\rm Kelvin}$), and $\Gamma$, of BB, NSA, and PL models, respectively.} \\
 \multicolumn{8}{p{13.5cm}}{$^c$ Effective source emitting region, calculated assuming a distance to M5 of 7.5 kpc. } \\
 \multicolumn{8}{p{13.5cm}}{$^d$ Unabsorbed fluxes in 0.5--2 keV and 2--8 keV energy bands, respectively, in units of 10$^{-16}$~erg~cm$^{-2}$~s$^{-1}$.} \\
 \multicolumn{8}{p{13.5cm}}{$^e$ Unabsorbed X-ray luminosities in 0.5--10 keV band, in units of 10$^{30}$~erg~s$^{-1}$.} \\
 \multicolumn{8}{p{13.5cm}}{$^f$ Log Bayesian evidences of fittings, normalized to the highest evidence in each case (i.e. 0 represents the model with highest likelihood). } \\
    \end{tabular}
    \label{tab:x-ray_fits}
\end{table*}

\section{Discussion}

\subsection{Properties of Pulsars in M5}
Previous to this study, we already knew that the pulsar population in M5 consists of fast-spinning pulsars (with the slowest object having a spin period of $\sim$8 ms) and is clearly dominated by binary systems: of the six previously known pulsars, five were in binaries, with a single isolated object, M5A. All the binaries have low-mass companions. All this is still true after our new binary discovery, M5G.

Our timing solutions provide precise positions for the pulsars. All of them are located within 1\farcm2 (2.8 core radii) of the centre of the GC, which is significantly smaller than the half-light and tidal radii ($\sim$1\farcm8 and $\sim$24\arcmin\, respectively, \citealt{Harris+2010}) that characterize the overall stellar distribution. This distribution is strongly reminiscent of that of 47~Tuc \citep{Heinke+2005}, which is the result of mass segregation (with the NSs, being more massive, migrating to the center of the GC via dynamical friction) followed by an equilibrium, or ``dynamical relaxation". This is typical of stellar populations with ages larger than the dynamical relaxation time, $\sim$ 0.2 Gyr for the core of M5 and $\sim 2.6$ Gyr for the average star in the cluster \citep{Harris+2010}.

Our timing results revealed weak surface magnetic fields (smaller than $6 \times 10^8$ G) and advanced characteristic ages (Gyrs,  Table~\ref{tab:accelerations}) for all seven pulsars in M5. For M5B, this can be additionally verified by its $\dot{P}_{\rm b}$. Another important indication about the nature of these systems is the low eccentricities of all the newly discovered binaries. The large eccentricity of M5B marks it as the only binary pulsar in M5 that has had substantial orbital perturbations by close stellar encounters. This is more likely for wider systems, which present a larger cross-section for such perturbations; and indeed M5B is precisely the binary with the largest orbital period. 

The precise positions determined from timing allowed optical detections of the companions of M5D, E and F (and, in an earlier work, of M5C). The colors and magnitudes of the companions point to a similar conclusion: they are very likely low-mass He WDs such as one would expect to find in low-eccentricity systems with fast MSPs. Their cooling ages are, again, of the order of a few Gyr.

Through the {\it Chandra} X-ray study of M5 MSPs, X-ray counterparts to four (MSPs C, D, E, and G) out of seven MSPs are identified, with new identifications of X-ray counterparts to MSPs D, E, and G, respectively. Thermal X-ray emission is observed from those four X-ray counterparts, whose spectra are well-described by an absorbed neutron star hydrogen atmosphere model. Their unabsorbed X-ray luminosities (0.5--10 keV) are between $\sim 6 \times 10^{30}$~erg~s$^{-1}$ and $\sim 1 \times 10^{31}$~erg~s$^{-1}$, which are in the range typically observed from thermally-emitting GC MSPs \citep{Zhao+2022}. The two black widows, MSPs C (eclipsing) and G (non-eclipsing), show little or no non-thermal X-ray emission, likely indicating weak synchrotron radiation produced by intra-binary shocks from these two spider pulsars. On the other hand, the X-ray luminosities of MSPs A, B, and F, which have no identified X-ray counterparts, are constrained to be lower than $\sim 4 \times 10^{30}$~erg~s$^{-1}$, while their X-ray counterparts could be detected in the future with deeper X-ray observations.

\subsection{The single-binary encounter rate}
In its broad parameters(predominance of binaries, fast spins, small B-fields and large characteristic ages, low orbital eccentricities, small companion masses, large cooling ages for a few of the He WD companions) the pulsar population of M5 is similar to the MSP population observed in the Galactic disk, as expected from the low $\gamma$ of M5. Thus, once an LMXB forms, it evolves undisturbed to form these types of systems, as in the Galactic disk.

The study of pulsars in low-$\gamma$ GCs like M5 provides an important test of the scenario presented by \cite{Verbunt_Freire2014}. If the apparently high-B pulsars in some GCs were being produced in large numbers by a process other than  LMXB disruption (for a summary of some of these processes, see section~\ref{sec:introduction} and, e.g., \citealt{Boyles_2011}) then there should be no reason why they cannot be produced in a low-$\gamma$ GC like M5. These alternative processes, like capture by a giant star and subsequent recycling, are, like the formation of low B-field MSPs, proportional to $\Gamma$. Therefore, the lack of high-B field pulsars relative to low-B field pulsars in low-$\gamma$ clusters like M5 - to which we add in this paper - adds to the evidence that the high-B field pulsars form in a process that depends on $\gamma$, like LMXB disruption. It is important to continue searching for high-B-field pulsars in low-$\gamma$ clusters; finding them might indicate alternative or additional formation channels for that type of pulsars.  

\subsection{The cluster dynamics}
The formation of MSPs is a combined effect of NS evolution and the cluster dynamics.
NSs were formed in the early stages of the GC because their progenitor massive stars have short lifetimes. A large fraction of NSs might escape from the GC due to high-velocity kicks after asymmetric supernovae. Over billions of years, the remaining NSs in the GC core experience almost all of the GC's dynamical evolution, including two-body relaxation driven mass segregation, core-collapse event(s) and the tidal dynamics between the GC and the Milky Way. All these mechanisms alter the density profile of GCs. The mass segregation leads to the central concentration of NSs, which then suffer less tidal effects from the Galaxy. The core collapse makes an ultra dense core, and leads to higher stellar encounter rates which can form pulsars, while the tidal dissolution stretches the cluster and strips stars away, regulating the stellar interaction rate.

The situation becomes more complex when considering the presence of a black hole (BH) population. Due to their significantly higher mass compared to NSs and normal stars, BHs tend to occupy the core region of the cluster until many of them are ejected through few-body interactions involving binary BHs and third bodies. As a result, the observed central surface density appears to be low, lacking the characteristic feature of core collapse \citep[e.g.][]{Breen2013}.
The existence of BHs might also reduce the stellar interaction rate between NSs and stars \citep{Ye2019}. Instead, NSs might interact with binary BHs and escape from the cluster core.

For the case of M5, no clear evidence has been found from its density profile \citep[see e.g.\ the GC parameter catalogs of][]{Baumgardt18, Miocchi+2013, Harris1996} \footnote{\url{https://people.smp.uq.edu.au/HolgerBaumgardt/globular/}}$^{;}$\footnote{2010 edition: \url{https://physics.mcmaster.ca/~harris/mwgc.dat}} or its distribution of blue straggler stars that the cluster has experienced core-collapse event(s) 
\citep[see the review paper][for using blue straggler stars as a ``dynamical clock'' to date the time of core collapse] {Ferraro2020}. 
Thus it is not expected that the inner-cluster dynamics can result in very high stellar encounter rates to form MSPs.

However, M5 is known to have at least one stellar tidal stream \citep[see e.g.][]{Mateu2023}, which provides evidence of significant mass loss due to tidal interactions with the gravitational potential of the Milky Way.  The trailing tidal tail of M5 spans an extensive portion of the sky, covering $\sim$ 50 $\deg$ \citep{Grillmair2019}. This suggests that M5 has lost a lot of mass at its outer boundary and should have been much heavier in its early life than it appears now. 
Given its likely origin as an accreted GC from outside the Milky Way \citep{Massari2019}, M5 has a highly eccentric orbit \citep[$e>0.8$, see e.g.][]{Vasiliev21}, with its orbital perigalacticon very close to the Galactic center   
\footnote{see the orbit integration figure of M5 on \url{https://people.smp.uq.edu.au/HolgerBaumgardt/globular/fits/ngc5904.html}}. Consequently, the strong tidal forces near the Galactic center, particularly the tidal shocks, are likely responsible for the significant loss of stars that occurs each time the cluster passes the perigalacticon \citep[see also discussions in][]{Grillmair2019}. This mass loss affects not only the outer region of the cluster, but also the cluster core region. 
Indeed, based on HST data and N-body simulation, \citet{Baumgardt18} have determined that the current mass of M5 is 3.72$\times$10$^5 M_{\odot}$, with a half-mass radius ($\rh$) of 5.58 parsec. However, M5 exhibits a relatively low central density ($4.68\times10^3$ $M_{\odot}/{\rm pc}^3$), and central escape velocity (30.4 km/s). While its half-mass radius is typical for a cluster of its mass and Galactocentric radius, the lower central density and central escape velocity of M5 suggest that its core has undergone significant stellar depletion, including the loss of NSs.

By neglecting the evolution of the Milky-Way potential, we can estimate the dissolution time ($\Tdis$) of M5, which represents the member star dispersion timescale of the cluster, using the formula derived by \cite{Baumgardt2003,Wang2020},
\begin{equation}
    \label{eq:tdis} 
    \Tdis \sim \Trh^{3/4}\Tcr^{1/4} (\rt/\rh)^{3/2} (1-e)
\end{equation}
where $\Trh$, $\Tcr$, $\rt$, and $e$ are the half-mass relaxation time, the crossing time, tidal radius, and eccentricity of the star cluster’s orbit around the Galaxy, respectively.

For M5, the present-day values are approximately 
 $\Trh\approx3.2$~Gyr, $\Tcr\approx0.32$~Myr and $\rt\approx81$~pc according to \cite{Baumgardt18}.
The corresponding dissolution time is $\Tdis\approx4.3$~Gyr, i.e. the timescale over which the cluster disperses a significant portion of its member stars is shorter than the cluster age \citep[$\sim$ 11.5 Gyr, ][]{VandenBerg2013}.  
Therefore, it can be inferred that M5 has experienced significant mass loss in the past. Moreover, considering the possibility of the existence of a BH population, the $\Tdis$ of M5 might be even shorter \citep{Wang2020}.
We contend that due to the significant mass loss experienced by M5, the potential ejection of NSs through few-body interactions within the core, and the initial natal kick imparted to NSs during supernova events, a substantial fraction of NSs, even MSPs, have been lost over the course of M5's history.

The present-day mass function of M5, as observed in \cite{Baumgardt2017}, exhibits a power-law index of $\alpha \approx -0.85$ within the mass range of $0.2$ to $0.8~M_\odot$. Referring to the \cite{Kroupa2001} initial mass function, which has an index of $\alpha \approx -1.3$, the higher $\alpha$ in M5 suggests that mass segregation may have occurred, leading to a preferential loss of low-mass stars through tidal evaporation. Therefore, we anticipate that the NS fraction on the tidal stream will be lower than that in the cluster.

The dynamical evolution of M5 discussed above shapes the cluster density profile and stellar mass function. Consequently, it also affects the NS population and the collision rate in the cluster.
The remaining MSPs and their distribution are the selection effect resulting from the dynamical evolution that M5 has undergone.

\section{Summary}
We have carried out a comprehensive multi-wavelength study of the relatively low-mass, likely extra-galactic accreted, non-core-collapsed globular cluster, M5. We utilized data from Arecibo, FAST, HST, Chandra, and Fermi. The main results are as follows: 
\begin{enumerate}[1.]
\item With FAST, we discovered PSR J1518+0204G, the 7th pulsar in M5. M5G, which is in a black widow binary system, has a $\sim$ 2.75\,ms spin period and 0.11\,day orbital period.

\item All seven pulsars are fast MSPs with spin period $P<8$ ms. Five out of seven pulsars in M5 are in low eccentricity binaries, with low mass companions.
The average orbital eccentricity of the M5 binary pulsars is $e=0.024$. The general GC pulsar population has an average spin period $P=18$\,ms and an average eccentricity $e=0.11$.

\item With data from both Arecibo and FAST, spanning 34 years, we achieved new phase-connected timing solutions for M5D, E, F, and G, and improved those for M5A, B, and C. These show that pulsars in M5 have relatively low B-fields ($< 6.2 \times 10^8$G) and large ($> 0.8$ Gyr) characteristic ages; they are likely similar to the MSP population in the Galactic disk. 

\item The proper motions of five pulsars (M5A to E) measured in this work, with an average value of $\mu_{\alpha}=4.14 \pm 0.14$\,mas yr$^{-1}$ and $\mu_{\delta}=-10.25 \pm 0.4$\,mas yr$^{-1}$, are consistent with the Gaia EDR3 proper motion of the cluster. All five pulsars are consistent with the escape velocity circle on the $\mu_{\alpha}$ -- $\mu_\delta$ plane.

\item We measured M5B's periastron advance rate as $\dot{\omega} = 0.01361(6)^\circ$. This represents an order-of-magnitude improvement over, while still within 1$\sigma$ of, previous measurements, resulting in updated mass measurements: $m_{\rm c} = 0.163^{+0.095}_{-0.020} \, \rm M_{\odot}$, $m_{\rm p} = 1.981^{+0.038}_{-0.088} \, \rm M_{\odot}$ and $M_{\rm T} = 2.157^{+0.028}_{-0.027} \, \rm M_{\odot}$ for the companion, pulsar, and the total mass of the M5B system. M5B remains a likely heavy neutron star, with little constraint on its inclination, except that it cannot be close to being edge on.

\item We detected the Shapiro delay in M5F, and possibly in M5D, in radio timing data. M5F was determined to possess a high inclination (close to $89^\circ$), $m_{\rm c} \, \sim \, 0.2 \, \rm M_{\odot}$ ($\sim \pm 50\%$ uncertainty), and $m_{\rm p} \, \sim \, 1.4 \, \rm M_{\odot}$. These values are consistent with those estimated based on optical data.

\item  The companions of M5D, E and F are detected on archival HST images, which shows that they are low-mass He WDs with cooling ages of a few Gyr. These are consistent with the large characteristic ages inferred from timing.

\item Pulsars C, D, E and G were also detected in Chandra data, from which X-ray spectra were newly extracted for M5D, E, and G and can be well-fitted by absorbed neutron star hydrogen atmosphere models. 

\item All characteristics of the pulsar population are consistent with the theoretical expectations for a low-$\gamma$ GC. The apparent lack of high-B pulsars relative to the low-B pulsars in M5 and other low-$\gamma$ clusters favors $\gamma$-dependent processes (such as LMXB disruption) for the formation of high B-field pulsars in GCs, while disfavoring $\Gamma$-dependent ones (such as giant-star-capture).

\item The evolutionary history and status of M5 is influenced by significant mass loss through tidal interaction with the Milky Way, as evidenced by its relatively low mass (3.72$\times10^{5}$ M$_\odot$), non-core-collapsed state, and having at least one stellar tidal stream. 
\end{enumerate}

\section*{Acknowledgments}
This work is supported by NSFC grant No. 11988101, 11725313, by the National Key R$\&$D Program of China No. 2017YFA0402600, by Key Research Project of Zhejiang Lab No. 2021PE0AC03, and by Chinese Academy of Sciences President’s International Fellowship Initiative. Grant No. 2023VMC0001.
L.Z is supported by ACAMAR Postdoctoral Fellowship and NSFC grant No. 12103069.
A.R. gratefully acknowledges support by the Chinese Academy of Sciences President’s International (PIFI) Fellowship Initiative. Grant No. 2023VMC0001 and by the National Astronomical Observatory of China (NAOC). 
J.C acknowledges support from the China Scholarship Council.
M.C and C.P acknowledge financial support from the project Light-on-Dark granted by MIUR through PRIN2017-2017K7REXT.  
S.D is the recipient of an Australian Research Council Discovery Early Career Award (DE210101738) funded by the Australian Government.
X.F thanks the support of the NSFC grant No. 12203100 and the China Manned Space Project with NO. CMS-CSST-2021-A08.
C.H is supported by NSERC Discovery Grant RGPIN-2016-04602.
X.H is supported by the NSFC grant No. 12041303.
S.M.R is a CIFAR Fellow and is supported by the NSF Physics Frontiers Center award 2020265.
J.Z. is supported by a China Scholarship Council scholarship No. 202108180023. 
The National Radio Astronomy Observatory is a facility of the National Science Foundation operated under cooperative agreement by Associated Universities, Inc" and also 
this work has been funded using resources from the INAF Large Grant 2022 “GCjewels” (P.I. Andrea Possenti) approved with the Presidential Decree 30/2022.  
Pulsar research at UBC is funded by an NSERC Discovery Grant and by the Canadian Institute for Advanced Research.
We thank Ramesh Karuppusamy and Matthew Bailes for a careful reading of the manuscript and helpful suggestions.

\bibliographystyle{aasjournal}
\bibliography{M5.bib}
\end{document}